\documentclass[reprint,amsmath,amssymb,aps,superscriptaddress]{revtex4-1}

\usepackage{mathtools}
\usepackage{graphicx}
\usepackage{dcolumn}
\usepackage{bm}
\usepackage{hyperref}
\usepackage{tikz}
\usepackage[italicdiff]{physics}

\begin{document}

\title{Tensor cross interpolation approach for quantum impurity problems based on the weak-coupling expansion}
\author{Shuta Matsuura}
\affiliation{Department of Physics, University of Tokyo, Hongo, Tokyo 113-0033, Japan}
\author{Hiroshi Shinaoka}
\affiliation{Department of Physics, Saitama University, Saitama 338-8570, Japan}
\author{Philipp Werner}
\affiliation{Department of Physics, University of Fribourg, 1700 Fribourg, Switzerland}
\author{Naoto Tsuji}
\affiliation{Department of Physics, University of Tokyo, Hongo, Tokyo 113-0033, Japan}
\affiliation{RIKEN Center for Emergent Matter Science (CEMS), Wako, Saitama 351-0198, Japan}

\date{\today}

\begin{abstract}
    We apply the tensor cross interpolation (TCI) algorithm to solve equilibrium quantum impurity problems with high precision based on the weak-coupling expansion.
    The TCI algorithm, a kind of active learning method, factorizes high-dimensional integrals that appear in the perturbative expansion into a product of low-dimensional ones, enabling us to evaluate higher-order terms efficiently.
    This method is free from the sign problem which quantum Monte Carlo methods sometimes suffer from, and allows one to directly calculate the free energy.
    We benchmark the TCI impurity solver on an exactly solvable impurity model, and find good agreement with the exact solutions. 
    We also incorporate the TCI impurity solver into the dynamical mean-field theory to solve the Hubbard model,
    and show that the metal-to-Mott insulator transition is correctly described with comparable accuracy to the Monte Carlo methods.
    Behind the effectiveness of the TCI approach for quantum impurity problems lies the fact that the integrands in the weak-coupling expansion naturally have a low-rank structure in the tensor-train representation.
\end{abstract}

\maketitle

\section{Introduction} \label{sec:intro}
    
    Solving quantum impurity problems is a fundamental issue in condensed matter physics. They describe localized impurity states hybridized with a bath of noninteracting particles, providing a basis for the understanding of the Kondo effect \cite{Kondo_effect1}, a prototypical phenomenon in quantum many-body physics. 
    The impurity model setup is realized not only in solid states \cite{Kondo_effect_in_solid_state1, Kondo_effect_in_solid_state2, Kondo_effect_in_solid_state3} 
    but also in artificial quantum systems such as quantum dots \cite{Kondo_effect_quantum_dot1,Kondo_effect_quantum_dot2,Kondo_effect_quantum_dot3} and ultracold atoms \cite{Kondo_effect_ultracold_atom1,Kondo_effect_ultracold_atom2,Kondo_effect_ultracold_atom3}.
    Quantum impurity problems also play a central role in the dynamical mean-field theory (DMFT) \cite{DMFT, DMFT_review, DMFT_Kotliar, noneq_DMFT_review} (or more generally, in quantum embedding methods \cite{Quantum_embedding}), which is a powerful tool to study strongly correlated systems in high dimensions. It has been applied to study the metal-to-Mott insulator transition \cite{MIT_Jarrell,MIT0,MIT2,MIT7}, high-temperature superconductors \cite{HTSC1, DCA_review, HTSC2, HTSC3} and disordered systems \cite{random1, random2, random3}.

    Various methods have been developed to solve quantum impurity problems,
    among which the continuous-time quantum Monte Carlo (CT-QMC) method \cite{CT-QMC,weak_coupling_expansion1,weak_coupling_expansion2,strong_coupling_expansion,CTQMC_comparison} is widely used as a versatile and numerically exact impurity solver. In CT-QMC, one expands the partition function and observables into infinite series, which can be classified into weak- and strong-coupling ones depending on the expansion parameter. Each term in the expansion (corresponding to a Feynman diagram) consists of high-dimensional integrals, which can be evaluated by stochastic sampling. While statistical errors can be controlled within the framework, there are several drawbacks: The errors scale as $1/\sqrt{N}$ ($N$ is the number of samples), so that the convergence is relatively slow (in the recently proposed quasi-Monte Carlo method the scaling is improved to $1/N$ \cite{quasi_QMC1,quasi_QMC2}). One may also encounter the infamous sign problem when the method is applied to multi-orbital models, multi-site clusters, spin-orbit coupled systems, and nonequilibrium impurity problems. It is also not straightforward to calculate the free energy using the CT-QMC solver.

    Some of these issues may be overcome by the recently developed tensor cross interpolation (TCI) algorithm \cite{TCI1, TCI2, TCI3, TCI4, TCI_noneq_weak_coupling, TCI_library}, in which multi-variable integrands to be integrated in a perturbative expansion are regarded as high-dimensional tensors. The TCI approximates such a high-dimensional tensor by a one-dimensional network of tensors called a tensor train or a matrix product state (MPS), which can be constructed from partial (selected) information of the original tensor through a certain ``interpolation'' scheme. This is to be contrasted to the singular value decomposition for constructing MPSs, which uses the full information of the tensor and becomes computationally demanding if the dimension of the tensor is large.
    The TCI is particularly useful when the high-dimensional tensor has a low-rank structure, for which an efficient (but often heuristic) search algorithm allows to find a quasi-optimal tensor-train representation.
    
    Once the low-rank tensor-train representation of the integrand is obtained, one can evaluate high-dimensional integrals efficiently by separately performing a series of one-dimensional integrations.
    Even if the integrand shows an oscillatory behavior with positive and negative contributions, the integration can be evaluated accurately with the TCI (as long as it has a low-rank structure), so that the calculation does not practically suffer from the sign problem.
    Furthermore, one can directly calculate the partition function and the free energy with the TCI approach. Previously, TCI has been used for equilibrium impurity problems based on the strong-coupling expansion \cite{TCI_strong_coupling}. There are also applications to nonequilibrium impurity problems using the weak-coupling expansion with a simple hybridization function \cite{TCI_noneq_weak_coupling} and the self-consistent strong-coupling expansion \cite{TCI_noneq_strong_coupling, TCI_noneq_strong_coupling2}.

    In this paper, we apply the TCI algorithm to equilibrium impurity problems formulated within the weak-coupling framework, which has not yet been explored. The weak-coupling approach has an advantage in studying cluster problems, for which the strong-coupling expansion is less suited due to the exponentially large dimension of the local Hilbert space.
    We first apply the TCI solver to an exactly solvable impurity model.
    The results show that the integrand in the weak-coupling expansion indeed has a low-rank structure, which allows us to calculate the expansion up to the 40th order. We find good agreement between the TCI and exact solutions.
    We also employ the TCI as an impurity solver for DMFT to solve the Hubbard model on the Bethe lattice, and compare the results with those of CT-QMC. While the method is based on the weak-coupling expansion, we find that the metal-to-Mott insulator crossover is well reproduced with a comparable accuracy to CT-QMC. The weak-coupling TCI solver can also be used to explore the first-order Mott transition slightly below the critical temperature where the metallic and insulating solutions coexist. We furthermore show results for the free energy of the lattice system, which is not easy to evaluate by CT-QMC.

    The paper is organized as follows. 
    Section \ref{sec:methods} provides an outline of the formalism. We review the weak-coupling expansion of the impurity problem, and explain how the TCI algorithm can be used to evaluate high-dimensional integrals.
    Section \ref{sec:results} presents the results of the weak-coupling TCI impurity solver.
    In Sec. \ref{subsec:exactly solvable model result}, we apply the TCI solver to the exactly solvable impurity model, and show that the TCI algorithm can efficiently evaluate high-order contributions.
    In Sec. \ref{subsec:DMFT result}, we solve the Hubbard model in the Mott transition (crossover) regime by incorporating the TCI solver with DMFT.
    The results given by the TCI solver agree with those of the quantum Monte Carlo method with high precision.
    Section \ref{sec:discussions} discusses the implication of the results and future perspectives of the weak-coupling TCI impurity solver.

\section{Method} \label{sec:methods}
\subsection{Weak-coupling expansion of the impurity problem} \label{subsec:weak-coupling}
    In this section, we review the weak-coupling expansion for the single-site impurity problem using the path-integral formalism.
    The effective action of the single-site impurity model is given by 
    \begin{subequations}
        \label{eq:SIAM action}
        \begin{align}
            S &= S_{0} + S_{\mathrm{int}}, \\
            S_{0} &= \int_{0}^{\beta} \dd{\tau} \sum_{\sigma} c_{\sigma}^{*}(\tau) \partial_{\tau} c_{\sigma}(\tau) \notag \\
            &\qquad+ \int_{0}^{\beta} \dd{\tau} \dd{\tau'} \sum_{\sigma} c_{\sigma}^{*}(\tau) \Delta_{\sigma}(\tau - \tau') c_{\sigma}(\tau'), \\
            S_{\mathrm{int}} &= \int_{0}^{\beta} \dd{\tau} U \qty(c_{\uparrow}^{*}(\tau) c_{\uparrow}(\tau) - \frac{1}{2}) \qty(c_{\downarrow}^{*}(\tau) c_{\downarrow}(\tau) - \frac{1}{2}),
        \end{align}
    \end{subequations}
    where $c_\sigma^\ast(\tau)$ is the creation operator of an electron with spin $\sigma$ at imaginary time $\tau$, $\beta$ is the inverse temperature, $\Delta_{\sigma}(\tau)$ denotes the hybridization function, which represents the coupling between the impurity and bath degrees of freedom, and $U$ is the interaction parameter.
    The chemical potential is set to $\mu = U/2$ to ensure that the system is half filled. 
    The extension to systems away from half filling is straightforward, but will not be discussed in this paper.
    Using this action, the partition function and the impurity Green's function can be expressed in the path-integral form as
    \begin{align}
        Z &= \int \mathcal{D}c^{*} \mathcal{D}c \,\, e^{-S},  
        \label{eq:partition function} \\
        G_{\sigma}(\tau) &= -\frac{1}{Z} \int \mathcal{D}c^{*} \mathcal{D}c \,\, c_{\sigma}(\tau) c_{\sigma}^{*}(0) e^{-S},
        \label{eq:Green's function}
    \end{align}
    respectively. 

    By expanding Eqs.~\eqref{eq:partition function} and \eqref{eq:Green's function} with respect to $U$ and applying Wick's theorem, we obtain formulae for the weak-coupling expansion \cite{weak_coupling_expansion1,CT-QMC}:
    \begin{align}
        &\frac{Z}{Z_0} = \sum_{n=0}^{\infty} (-U)^n \int_{S_{n}^{0,\beta}} \dd{\tau_1} \cdots \dd{\tau_n} (\det \bm{D}_{n}^{\uparrow}) (\det \bm{D}_{n}^{\downarrow}), \label{eq:expansion of partition function} \\
        &G_{\sigma}(\tau) = \frac{Z_{0}}{Z} \sum_{n=0}^{\infty} (-U)^n  \int_{S_{n}^{0,\beta}} \dd{\tau_1} \cdots \dd{\tau_n} \notag \\ &\hspace{45.0mm} \times(\det \tilde{\bm{D}}_{n}^{\sigma}) (\det \bm{D}_{n}^{\bar{\sigma}}), \label{eq:expansion of Green's function}
    \end{align}
    where $\bm{D}_{n}^{\sigma}$ and $\tilde{\bm{D}}_{n}^{\sigma}$ are $n\times n$ and $(n+1)\times (n+1)$ matrices defined by
    \begin{align}
        &\bm{D}_{n}^{\sigma} = 
        \begin{pmatrix}
            \mathcal{G}_{\sigma}(0^{-}) - 1/2 & \cdots & \mathcal{G}_{\sigma}(\tau_{1} - \tau_{n}) \\
            \vdots & \ddots & \vdots \\
            \mathcal{G}_{\sigma}(\tau_{n} - \tau_{1}) & \cdots & \mathcal{G}_{\sigma}(0^{-}) - 1/2
        \end{pmatrix}, \\
        &\tilde{\bm{D}}_{n}^{\sigma} =
        \begin{pmatrix}
            \mathcal{G}_{\sigma}(\tau) & \mathcal{G}_{\sigma}(\tau - \tau_{1}) & \cdots & \mathcal{G}_{\sigma}(\tau - \tau_{n}) \\
            \mathcal{G}_{\sigma}(\tau_{1}) & \mathcal{G}_{\sigma}(0^{-}) - 1/2 & \cdots &
            \mathcal{G}_{\sigma}(\tau_{1} - \tau_{n}) \\
            \vdots & \vdots & \ddots & \vdots \\
            \mathcal{G}_{\sigma}(\tau_{n}) & \mathcal{G}_{\sigma}(\tau_{n} - \tau_{1}) & \cdots & \mathcal{G}_{\sigma}(0^{-}) - 1/2
        \end{pmatrix}, \label{eq:Green's function determinant} \\
        &\mathcal{G}_{\sigma}(\tau) = - (\partial_{\tau} + \Delta_{\sigma}(\tau))^{-1}, 
        \label{eq:Weiss field}
    \end{align}
    $\mathcal G_\sigma(\tau)$ is the Weiss field,
    $Z_{0}$ is the partition function for the impurity model described by the noninteracting action $S_{0}$, and
    $S_{n}^{0,\beta}$ denotes a simplex defined by
    \begin{equation}
        S_{n}^{a,b} = \qty{(\tau_{1}, \cdots, \tau_{n}) \in \mathbb{R}^n \, | \, a \le \tau_{1} \le \cdots \le \tau_{n} \le b}.
    \end{equation}
    The index $\bar{\sigma}$ represents the spin polarization opposite to $\sigma$.

    By truncating the infinite series \eqref{eq:expansion of partition function} and \eqref{eq:expansion of Green's function} at a finite order $n_{\mathrm{max}}$, we obtain the following approximate expressions for the partition function and the Green's function:
    \begin{align}
        &\frac{Z}{Z_0} \simeq \sum_{n=0}^{n_{\mathrm{max}}} (-U)^n \int_{S_{n}^{0,\beta}} \dd{\tau_1} \cdots \dd{\tau_n} (\det \bm{D}_{n}^{\uparrow}) (\det \bm{D}_{n}^{\downarrow}), \label{eq:expansion of partition function finite order} \\
        &G_{\sigma}(\tau) \simeq \frac{Z_{0}}{Z} \sum_{n=0}^{n_\mathrm{max}} (-U)^n  \int_{S_{n}^{0,\beta}} \dd{\tau_1} \cdots \dd{\tau_n} \notag \\ &\hspace{45.0mm} \times(\det \tilde{\bm{D}}_{n}^{\sigma}) (\det \bm{D}_{n}^{\bar{\sigma}}). \label{eq:expansion of Green's function finite order}
    \end{align}
    
\subsection{Dynamical mean-field theory} \label{subsec:DMFT}
    Impurity problems play an important role in the dynamical mean-field theory (DMFT), which maps the lattice model to an effective impurity model. In this mapping, the hybridization function is self-consistently determined under the assumption that the lattice self-energy be local in space \cite{DMFT_review}.
    Solving the impurity problem is a key step in DMFT, enabling us to compute the local Green's function of the strongly correlated lattice system in high spatial dimensions.
    This section briefly reviews the formulation of DMFT for a simple case.

    Let us consider the Hubbard model on the Bethe lattice with connectivity $z$ at half filling ($\mu = U/2$).
    Its Hamiltonian is given by
    \begin{align}
        H &= -\frac{t}{\sqrt{z}} \sum_{\ev{i,j},\sigma} (c_{i\sigma}^{\dag} c_{j\sigma}+\mbox{H.c.})
        \notag
        \\
        &\quad + \sum_{i} U \qty(n_{i\uparrow} - \frac{1}{2}) \qty(n_{i\downarrow} - \frac{1}{2}),
        \label{eq:Hubbard model Bethe lattice}
    \end{align}
    where $t$ is the nearest-neighbor hopping amplitude.
    The local Green's function for the lattice system with the Hamiltonian \eqref{eq:Hubbard model Bethe lattice} is defined by
    \begin{equation}
        G_{\mathrm{loc}}(\tau) = - \ev{T_{\tau} c_{i\sigma}(\tau) c_{i\sigma}^{\dag}(0)},
    \end{equation}
    where $T_\tau$ is the time-ordering operator.
    The local lattice Green's function is identical to that of the single-site impurity model [Eq. \eqref{eq:SIAM action}] with
    \begin{equation}
        \Delta_{\sigma}(\tau) = t^2 G_{\mathrm{loc}}(\tau),
        \label{eq:new hybridization function}
    \end{equation}
    when we consider the infinite coordination limit $z \rightarrow \infty$ \cite{DMFT_review, infinite_dim} (with bandwidth $4t$).
    Here, we assume that the system is in the paramagnetic phase, where the local Green's function does not depend on spin.

    Note that $G_{\mathrm{loc}}(\tau)$ to be computed is included in the definition of the impurity problem [Eq.~\eqref{eq:expansion of Green's function}]. 
    $G_{\mathrm{loc}}(\tau)$ must be determined self-consistently by iterating the following loop until convergence is reached:
    Starting from an initial guess of the Weiss field $\mathcal{G}(\tau)$, (i) compute $G(\tau)$ by Eqs.~\eqref{eq:expansion of partition function finite order} and \eqref{eq:expansion of Green's function finite order}, 
    (ii) obtain $\Delta_{\sigma}(\tau)$ from Eq.~\eqref{eq:new hybridization function} by setting $G_{\mathrm{loc}}(\tau) = G(\tau)$, 
    (iii) solve Eq.~\eqref{eq:Weiss field} to update $\mathcal{G}(\tau)$.
\subsection{TCI algorithm}
    \label{subsec:TCI}
    In this section, we review the tensor cross interpolation (TCI) and how it can be used to evaluate high-dimensional integrals based on Ref.~\cite{TCI_library}.
    The TCI approximates a tensor by a product of its low-dimensional slices, which is called a tensor train.
    Here, we mainly focus on the definition of the TCI approximation and its key properties.
    We briefly describe the algorithm used to construct the tensor train, for the details of which we refer to the paper of Y. Núñez-Fernández \textit{et al.} \cite{TCI_library}.

    \subsubsection{Matrix cross interpolation}
    We first introduce the matrix cross interpolation (CI) \cite{TCI_library,TCI1,TCI2,TCI3,TCI4,TCI_noneq_weak_coupling}, which forms a basis of the TCI approximation.
    Let us consider an $M \times N$ matrix $A$.
    We define the set of the row and column indices of $A$ as 
    $\mathbb{I} = \qty{1, 2, \cdots, M}$ and $\mathbb{J} = \qty{1, 2, \cdots, N}$, respectively, and write the subset of $\mathbb{I}$, $\mathbb{J}$ with size $\chi \, (\le \rank A)$ as 
    $I = \qty{i_{1}, i_{2}, \cdots, i_{\chi}} \subset \mathbb{I}$ and 
    $J = \qty{j_{1}, j_{2}, \cdots, j_{\chi}} \subset \mathbb{J}$.
    The $\chi \times \chi$ submatrix (or slices) of $A$ which consists of rows and columns with indices in $I$, $J$ is denoted by $A(I, J)$, i.e.,
    \begin{equation}
        \qty[A(I, J)]_{\alpha \beta} = A_{i_{\alpha} j_{\beta}}.
    \end{equation}
    In a similar manner, an $M \times \chi$ submatrix $A(\mathbb{I}, J)$ and a $\chi \times N$ submatrix $A(I, \mathbb{J})$ are defined.
    With these notations, the CI formula approximates the matrix $A$ by a rank-$\chi$ matrix $\tilde{A}$ defined as
    \begin{equation}
        A \simeq A(\mathbb{I},J) A(I, J)^{-1} A(I, \mathbb{J}) \eqqcolon \tilde{A}.
        \label{eq:CI formula}
    \end{equation}
    The matrix $A(I, J)$ is called a pivot matrix, and its elements are called pivots.
    Note that the data size is compressed from $\order{MN}$ to $\order{\max \qty{M,N}\chi}$ by the CI formula when $A$ is approximated with a small number of pivots $(\chi \ll M,N)$.
    
    The matrix $\tilde{A}$ has two important properties that ensure the validity of the approximation in Eq.~\eqref{eq:CI formula}.
    The first one is
    \begin{equation}
        A(\mathbb{I}, J) = \tilde{A}(\mathbb{I}, J), \quad
        A(I, \mathbb{J}) = \tilde{A}(I, \mathbb{J}),
        \label{eq:CI property}
    \end{equation}
    which means that the approximation reproduces the original matrix $A$ in the selected rows $I$ and columns $J$, and hence
    all the other elements, not included in $I$ or $J$, are interpolated.
    The second one is that if $\chi$ is equal to the rank of the matrix $A$, the approximation becomes exact, i.e., $A = \tilde{A}$.

    To achieve a high-quality approximation, $I$ and $J$ should be chosen carefully.
    Although finding the optimal pivots is computationally demanding, a heuristic search algorithm based on the partial rank-revealing LU (prrLU) decomposition is known to find suboptimal ones \cite{TCI_library}.
    By repeating the process of searching for the matrix element with largest modulus and performing the Gauss elimination $\chi$ times, one can construct the rank-$\chi$ CI approximation of a given matrix.

    \subsubsection{Tensor cross interpolation}
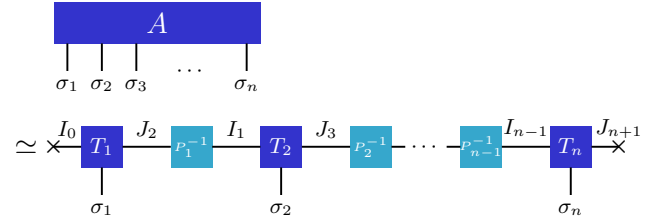
\begin{figure}[t]
        \centering
        \resizebox{\columnwidth}{!}{%
        \begin{tikzpicture}[thick]

            \fill[blue!60!gray] (-0.2, 4.0) rectangle (2.8, 4.6);
            \draw[white] (1.3, 4.3) node {\large{$A$}};
            \draw (0.0, 4.0) -- ++(0, -0.4) node[below]{$\sigma_{1}$};
            \draw (0.5, 4.0) -- ++(0, -0.4) node[below]{$\sigma_{2}$};
            \draw (1.0, 4.0) -- ++(0, -0.4) node[below]{$\sigma_{3}$};
            \draw (2.6, 4.0) -- ++(0, -0.4) node[below]{$\sigma_{n}$};
            \draw (1.8, 3.6) node{$\cdots$};

            \draw (-0.6, 2.5) node{\large{$\simeq$}};

            \fill[blue!60!gray] (0.2, 2.2) rectangle (0.8, 2.8);
            \draw[white] (0.5, 2.5) node {$T_{1}$};
            \fill[cyan!60!gray] (1.5, 2.2) rectangle (2.1, 2.8);
            \draw[white] (1.8, 2.5) node {\tiny{$P_{1}^{-1}$}};
            \fill[blue!60!gray] (2.8, 2.2) rectangle (3.4, 2.8);
            \draw[white] (3.1, 2.5) node {$T_{2}$};
            \fill[cyan!60!gray] (4.1, 2.2) rectangle (4.7, 2.8);
            \draw[white] (4.4, 2.5) node {\tiny{$P_{2}^{-1}$}};
            \fill[cyan!60!gray] (5.7, 2.2) rectangle (6.3, 2.8);
            \draw[white] (6.0, 2.5) node {\tiny{$P_{n-1}^{-1}$}};
            \fill[blue!60!gray] (7.0, 2.2) rectangle (7.6, 2.8);
            \draw[white] (7.3, 2.5) node {$T_{n}$};

            \draw (0.5, 2.2) -- ++(0, -0.4) node[below]{$\sigma_{1}$};
            \draw (3.1, 2.2) -- ++(0, -0.4) node[below]{$\sigma_{2}$};
            \draw (7.3, 2.2) -- ++(0, -0.4) node[below]{$\sigma_{n}$};

            \draw (-0.2, 2.5)node{$\bm{\times}$} --node[midway,above]{$I_{0}$} (0.2, 2.5);
            \draw (0.8, 2.5) --node[midway,above]{$J_{2}$} (1.5, 2.5);
            \draw (2.1, 2.5) --node[midway,above]{$I_{1}$} (2.8, 2.5);
            \draw (3.4, 2.5) --node[midway,above]{$J_{3}$} (4.1, 2.5);
            \draw (4.7, 2.5) -- (4.9, 2.5);
            \draw (5.5, 2.5) -- (5.7, 2.5);
            \draw (6.3, 2.5) --node[midway,above]{$I_{n-1}$} (7.0, 2.5);
            \draw (7.6, 2.5) -- (8.0, 2.5)node{$\bm{\times}$};
            \draw (8.0, 2.5) node[above]{$J_{n+1}$};

            \draw (5.2, 2.5) node{$\cdots$};
        \end{tikzpicture}
        }
        \caption{Tensor-train decomposition of a tensor $A_{\sigma_{1} \cdots \sigma_{n}}$ in the TCI. 
        $P_{\ell}$ and $T_{\ell}$ represent zero-dimensional and one-dimensional slices of the tensor $A$ defined by Eqs.~\eqref{eq:def of P tensor} and \eqref{eq:def of T tensor}, respectively.
        By decomposing $A_{\sigma_{1} \cdots \sigma_{n}}$ into a product of small tensors, one can perform high-dimensional integrals efficiently.}
        \label{fig:tensor train}
    \end{figure}

    The tensor cross interpolation (TCI) is an extension of the matrix CI that approximates a tensor by a product of its low-dimensional slices \cite{TCI_library,TCI1,TCI2,TCI3,TCI4,TCI_noneq_weak_coupling}.
    Let us consider a tensor $A$ with $n$ legs labeled by $\sigma_{1}, \sigma_{2}, \cdots, \sigma_{n}$.
    We assume that the index $\sigma_{\ell} \, (1 \le \ell \le n)$ takes values from the set $\mathbb{S}_{\ell} = \qty{1, 2, \cdots, d_{\ell}}$, which means that
    $A$ is a $d_{1} \times d_{2} \times \cdots \times d_{n}$ tensor.
    Let us define the set of the row multi-indices $\mathbb{I}_{\ell}$ and column multi-indices $\mathbb{J}_{\ell}$ for $1 \le \ell \le n$ by
    \begin{align}
        \mathbb{I}_{\ell} &= \mathbb{S}_{1} \times \mathbb{S}_{2} \times \cdots \times \mathbb{S}_{\ell}, \\
        \mathbb{J}_{\ell} &= \mathbb{S}_{\ell} \times \mathbb{S}_{\ell+1} \times \cdots \times \mathbb{S}_{n},
    \end{align}
    and their subsets by $I_{\ell} \subset \mathbb{I}_{\ell}$ and $J_{\ell} \subset \mathbb{J}_{\ell}$.
    For convenience, we define $I_{0}$ and $J_{n+1}$ as a set which consists of an empty tuple,
    \begin{equation}
        I_{0} = J_{n+1} = \qty{()}.
    \end{equation}
    We require the number of the elements of the subset $I_{\ell}$ and $J_{\ell+1}$ to be equal to each other for $1 \le \ell \le n-1$, 
    \begin{equation}
        \abs{I_{\ell}} = \abs{J_{\ell+1}} \eqqcolon \chi_{\ell} \quad (1 \le \ell \le n-1).
    \end{equation} 
    For $\ell = 0$ and $n$, we define $\chi_{0} = \abs{I_{0}} = 1$ and $\chi_{n} = \abs{J_{n+1}} = 1$.
    The slices of the tensor $A$ are defined in a similar manner to the matrix case, e.g.,
    $A(I_{\ell}, J_{\ell+1})$ is a $\chi_{\ell} \times \chi_{\ell}$ matrix, and
    $A(I_{\ell-1}, \mathbb{S}_{\ell}, J_{\ell+1})$ is a $\chi_{\ell-1} \times d_{\ell} \times \chi_{\ell}$ three-leg tensor.
    Using these notations, the TCI approximation $\tilde A$ of $A$ is defined as
    \begin{align}
        A_{\sigma_{1} \cdots \sigma_{n}} &\simeq T_{1}^{\sigma_{1}} P_{1}^{-1} T_{2}^{\sigma_{2}} P_{2}^{-1} \cdots P_{n-1}^{-1} T_{n}^{\sigma_{n}} \notag \\
        &\eqqcolon \tilde{A}_{\sigma_{1} \cdots \sigma_{n}}.
    \end{align}
    Here, $P_{\ell}$ is a $\chi_{\ell} \times \chi_{\ell}$ matrix defined by
    \begin{equation}
        P_{\ell} = A(I_{\ell}, J_{\ell+1}),
        \label{eq:def of P tensor}
    \end{equation}
    $T_{\ell}$ is a $\chi_{\ell-1} \times d_{\ell} \times \chi_{\ell}$ three-leg tensor defined by
    \begin{equation}
        T_{\ell} = A(I_{\ell-1}, \mathbb{S}_{\ell}, J_{\ell+1}),
        \label{eq:def of T tensor}
    \end{equation}
    and $T_{\ell}^{\sigma_{\ell}}$ is a $\chi_{\ell-1} \times \chi_{\ell}$ matrix whose elements are related to those of $T_{\ell}$ by
    \begin{equation}
    \qty[T_{\ell}^{\sigma_{\ell}}]_{ij} = \qty[T_{\ell}]_{i \sigma_{\ell} j}.
        \label{eq:def of T matrix}
    \end{equation}
    The matrix $P_{\ell}$ is called a pivot matrix, and $\chi = \max_{\ell} \chi_{\ell}$ is the maximum bond dimension or the rank of the tensor train.
    This decomposition is diagrammatically represented in Fig.~\ref{fig:tensor train}.

    The TCI approximation has a property similar to Eq.~\eqref{eq:CI property}.
    To state this property, we introduce the nesting condition.
    If $I_{\ell} \subset I_{\ell-1} \times \mathbb{S}_{\ell}$, 
    $I_{\ell}$ is said to be nested with respect to $I_{\ell-1}$, and we write $I_{\ell-1} < I_{\ell}$.
    In the same way, if $J_{\ell} \subset \mathbb{S}_{\ell} \times J_{\ell+1}$, 
    $J_{\ell}$ is said to be nested with respect to $J_{\ell+1}$, and we write $J_{\ell} > J_{\ell+1}$.
    When $I_{0} < I_{1} < \cdots < I_{\ell-1}$ and $J_{\ell+1} > J_{\ell+2} > \cdots > J_{n}$ holds, $\tilde{A}$ satisfies
    \begin{equation}
        A(I_{\ell-1}, \mathbb{S}_{\ell}, J_{\ell+1}) = \tilde{A}(I_{\ell-1}, \mathbb{S}_{\ell}, J_{\ell+1}).
    \end{equation}
    This means that the TCI approximation can reproduce some of the elements of the original tensor, and the other elements are interpolated if the nesting condition is satisfied.

    As is the case in the matrix CI, the quality of the TCI approximation heavily depends on the choice of $I_{\ell}$ and $J_{\ell}$.
    To find the suboptimal $I_{\ell}$ and $J_{\ell}$, the two-site TCI algorithm \cite{TCI_library} is available.
    In this algorithm, the matrix CI with prrLU is repeatedly applied to the two-dimensional slices of the tensor $A$ to construct a tensor train.
    If a tensor $A$ has a structure that can be approximated with a low-rank tensor train, 
    it is empirically known that this algorithm can almost always construct such a low-rank tensor train.
    One of the notable features of this algorithm is that it does not require the full information of the tensor $A$ to construct the tensor-train representation.
    This is in contrast to the singular value decomposition, which requires the full information of the tensor and is less efficient for decomposing a high-dimensional tensor.
    Since the TCI algorithm actively samples a few important elements during the construction of the tensor train, it can be regarded as a kind of active learning algorithm.

    \subsubsection{Evaluating high-dimensional integrals by TCI}
    To evaluate the Green's function of the impurity model, we have to perform the high-dimensional integrals appearing in Eqs.~\eqref{eq:expansion of partition function finite order} and \eqref{eq:expansion of Green's function finite order}.
    In this work, we evaluate these integrals by combining Gauss--Kronrod (GK) quadrature and the tensor-train decomposition generated by the TCI algorithm following the approach in Ref.~\cite{TCI_noneq_weak_coupling,TCI_library}.
    This section reviews how the high-dimensional integrals can be performed with the GK quadrature and the TCI algorithm.

    Let us consider an $n$-dimensional integral of a continuous function $f(x_{1}, \cdots, x_{n})$ over a hypercube, 
    \begin{equation}
        \mathcal{I} = \int_{[0,1]^{n}} \dd[n]{\bm{x}} f(\bm{x}).
    \end{equation}
    By using the GK quadrature rule, the integral can be approximated as a discrete sum,
    \begin{align}
        \mathcal{I} &\simeq \sum_{\sigma_{1}=1}^{d} \cdots \sum_{\sigma_{n}=1}^{d} w_{\sigma_{1}} \cdots w_{\sigma_{n}} f(x_{\sigma_{1}}, \cdots, x_{\sigma_{n}}) \notag \\
        &\eqqcolon \sum_{\sigma_{1}=1}^{d} \cdots \sum_{\sigma_{n}=1}^{d} A_{\sigma_{1} \cdots \sigma_{n}},
    \end{align}
    where $x_{i}$ are the zeros of the Legendre polynomials and Stieltjes polynomials, $w_{i}$ are the corresponding weights, and
    $d$ denotes the order of the GK quadrature which is set to $d = 15$ in this work.
    (The convergence with respect to $d$ is fast in the problem considered in Sec.~\ref{sec:results}, so using $d$ larger than $15$ makes little difference in the accuracy.)
    The computational cost of this $n$-fold sum is $\order{d^{n}}$, which grows exponentially with respect to the integral dimension $n$.
    Note that the weights $w_{\sigma_{\ell}}$ are absorbed in the definition of the tensor $A$ (although it is also possible to formulate the method without absorbing them).
    
    To reduce the computational cost, we use the TCI algorithm  \cite{TCI1, TCI4, TCI_library}.
    In the TCI approach, the tensor $A_{i_{1} \cdots i_{n}}$ is decomposed into a product of small tensors,
    \begin{align}
        A_{\sigma_{1} \cdots \sigma_{n}} &\simeq T_{1}^{\sigma_{1}} P_{1}^{-1} T_{2}^{\sigma_{2}} P_{2}^{-1} \cdots P_{n-1}^{-1} T_{n}^{\sigma_{n}} \notag \\
        &= M_{1}^{\sigma_{1}} M_{2}^{\sigma_{2}} \cdots M_{n}^{\sigma_{n}}
    \end{align}
    by sampling few elements of $A_{i_{1} \cdots i_{n}}$,
    where $P_{\ell}$ and $T_{\ell}^{\sigma_{\ell}}$ are defined by Eqs.~\eqref{eq:def of P tensor}--\eqref{eq:def of T matrix} and $M_{\ell}^{\sigma_{\ell}}$ is defined by
    \begin{equation}
        M_{\ell}^{\sigma_{\ell}} = 
        \begin{cases}
            T_{\ell}^{\sigma_{\ell}} P_{\ell}^{-1} & \text{if $1 \le \ell \le n-1$,} \\
            T_{n}^{\sigma_{n}} & \text{if $\ell = n$.}
        \end{cases}
    \end{equation}
    By this decomposition, the original $n$-fold summation can be reduced to $n$ independent summations,
    \begin{equation}
        \mathcal{I} \simeq \qty(\sum_{\sigma_{1}=1}^{d} M_{1}^{\sigma_{1}}) \qty(\sum_{\sigma_{2}=1}^{d} M_{2}^{\sigma_{2}}) \cdots \qty(\sum_{\sigma_{n}=1}^{d} M_{n}^{\sigma_{n}}).
    \end{equation}
    The resulting computational cost is $\order{nd\chi^{2}} \ll \order{d^{n}}$, where $\chi$ is the maximum bond dimension of the tensor-train decomposition.

    To implement the TCI algorithm, we utilize \texttt{TensorCrossInterpolation.jl} introduced in Ref.~\cite{TCI_library}.
    It provides the TCI algorithm based on the prrLU decomposition, and allows one to construct the tensor-train representation without inverting 
    the pivot matrix $P_{\ell}$, which can be ill-conditioned.
    Also, a global search algorithm to find pivots is available, which helps to avoid being stuck in a subpart of the whole space while sampling the elements of the tensor.
    These features have not been incorporated in the previous works~\cite{TCI_noneq_weak_coupling, TCI_strong_coupling}.
    While the algorithm described above does not preserve the nesting condition by default, practically no problem has been observed in the numerical calculations \cite{TCI_library}.
    
\subsection{Computation of the partition function}  \label{subsec:computation of partition function}

    In this section, we discuss the technical aspects of performing the integrals that appear in the expression of the partition function [Eq.~\eqref{eq:expansion of partition function finite order}]. 
    The technique explained in this section and the following section is based on the paper of A. Erpenbeck \textit{et al.} \cite{TCI_strong_coupling}.
    The map employed in this section is described in detail in Appendix \ref{app:change var}.
    
    In order to perform high-dimensional integrals based on the method described in Sec.~\ref{subsec:TCI}, the integral domain must be a hypercube.
    Since the integral in Eq.~\eqref{eq:expansion of partition function finite order},
    \begin{equation}
        \mathcal{I}_{n} = (-U)^{n} \int_{S_{n}^{0,\beta}} \dd{\tau_{1}} \cdots \dd{\tau_{n}} P(\tau_{1}, \cdots, \tau_{n}),
        \label{eq:partition function integral}
    \end{equation}
    where $P(\tau_{1}, \cdots, \tau_{n}) = (\det \bm{D}_{n}^{\uparrow}) (\det \bm{D}_{n}^{\downarrow})$, is defined on the simplex $S_{n}^{0,\beta}$, one has to change the integral variables to rewrite the integral over a simplex in the form of an integral over a hypercube $[0,1]^{n}$.
    To this end, we use a bijective map, $h_{0,\beta} : [0,1]^{n} \rightarrow S_{n}^{0,\beta}$, which is defined in Eqs.~\eqref{eq:map from normalized simplex to simplex}, \eqref{eq:map from hypercube to normalized simplex} and \eqref{eq:map from hypercube to simplex}.
    By relating $(v_{1}, \cdots, v_{n}) \in [0,1]^{n}$ with $(\tau_{1}, \cdots, \tau_{n}) \in S_{n}^{0,\beta}$ through
    \begin{equation}
        (\tau_{1}, \cdots, \tau_{n}) = h_{n}^{0,\beta}(v_{1}, \cdots, v_{n}),
        \label{eq:partition function change variable}
    \end{equation}
   the integral $\mathcal{I}_{n}$ can be expressed as
   \begin{multline}
       \mathcal{I}_{n} = (-U)^{n} \int_{[0,1]^{n}} \dd{v_{1}} \cdots \dd{v_{n}} P(h_{n}^{0,\beta}(v_{1}, \cdots, v_{n})) \\ \times J_{h_{n}^{0,\beta}}(v_{1}, \cdots, v_{n}),
   \end{multline}
   where $J_{h_{0,\beta}}(v_{1}, \cdots, v_{n})$ is the Jacobian for the transformation \eqref{eq:partition function change variable} explicitly given by Eq.~\eqref{eq:Jacobian}.
   Let us remark here that this Jacobian is separable, i.e., can be factorized as
   \begin{equation}
       J_{h_{n}^{0,\beta}}(v_{1}, \cdots, v_{n}) = A_{1}(v_{1}) \cdots A_{n}(v_{n}),
   \end{equation}
   which means that it has a low-rank structure.
   
   For later convenience, we define the function
   \begin{equation}
       \tilde{P}(v_{1}, \cdots, v_{n}) = P(h_{n}^{0,\beta}(v_{1}, \cdots, v_{n})) J_{h_{n}^{0,\beta}}(v_{1}, \cdots, v_{n}), \label{eq:partition integrand without Jacobian}
   \end{equation}
   and write $\mathcal{I}_{n}$ as 
   \begin{equation}
       \mathcal{I}_{n} = (-U)^{n} \int_{[0,1]^{n}} \dd{v_{1}} \cdots \dd{v_{n}} \tilde{P}(v_{1}, \cdots, v_{n}).
   \end{equation}   
   We may then express the partition function as
   \begin{equation}
       \frac{Z}{Z_{0}} \simeq \sum_{n=1}^{n_{\mathrm{max}}} (-U)^{n} \int_{[0,1]^{n}} \dd{v_{1}} \cdots \dd{v_{n}} \tilde{P}(v_{1}, \cdots, v_{n}).
       \label{eq:partition integrand on the hypercube}
   \end{equation}

   To calculate the integral in Eq.~\eqref{eq:partition integrand on the hypercube} by the TCI approach, the order of the legs has to be specified. 
   The most natural choice is $(v_{1}, \cdots, v_{n})$, but since the integrand in Eq.~\eqref{eq:partition integrand on the hypercube} is not invariant under the permutation of $v_{1}, \cdots, v_{n}$, the efficiency of the calculation could be improved by reordering the legs.
   To investigate the effect of the order, we calculated the integral with three different orderings, $(v_{1}, \cdots, v_{n})$, $(v_{\lfloor n/2+1 \rfloor}, v_{2}, \cdots, v_{1}, \cdots, v_{n})$ and $(v_{n}, v_{2}, \cdots, v_{1})$, and found that there is no significant change in the speed of convergence with respect to the bond dimension.
   Thus, we use the order $(v_{1}, \cdots, v_{n})$ in the following.

   \subsection{Computation of the Green's function} \label{subsec:computation of Green's function}
    In this section, we discuss how to apply the TCI algorithm to the integral in Eq.~\eqref{eq:expansion of Green's function finite order},
    \begin{equation}
        \mathcal{J}_{n}^{\sigma}(\tau) = (-U)^{n} \int_{S_{n}^{0,\beta}} \dd{\tau_{1}} \cdots \dd{\tau_{n}} Q^{\sigma}(\tau_{1}, \cdots, \tau_{n}; \tau), \label{eq:Green's function integral}
    \end{equation}
    where $Q^{\sigma}(\tau_{1}, \cdots, \tau_{n}; \tau) = (\det \tilde{\bm{D}}_{n}^{\sigma}) (\det \bm{D}_{n}^{\bar{\sigma}})$.
    Similarly to the case of the partition function, we have to change the integral variables so that the integral domain becomes a hypercube. However, the situation is more complicated due to the existence of a non-integrated variable ($\tau$) and the discontinuity of the integrand. 
    We will explain how to deal with these issues in the following subsections.
    
    \subsubsection{Treatment of the non-integrated variable \texorpdfstring{$\tau$}{Lg}}
    Unlike the integral appearing in the partition function [Eq.~\eqref{eq:partition function integral}], the one in the Green's function [Eq.~\eqref{eq:Green's function integral}] involves the variable $\tau$, which is not integrated.

    There are (at least) two ways to deal with it. The first one is to get the tensor-train approximation of $Q^{\sigma}(\tau_{1}, \cdots, \tau_{n}; \tau)$ at each sampling point $\tau$, and calculate $\mathcal{J}_{n}^{\sigma}(\tau)$. 
    Although this method is stable, the computational cost is proportional to 
    the number of sampling points $\tau$, which scales as $\order{\ln \omega_\mathrm{max}\beta}$ when using the intermediate representation basis \cite{sparseIR1, sparseIR2} ($\omega_\mathrm{max}$ is an ultraviolet cutoff on the real-frequency axis).

    Another method is to treat $\tau$ as a leg of the tensor, obtain a tensor-train representation for this enlarged tensor, and then take a summation over the legs other than $\tau$.
    In this method, we can get the Green's function for all the sampling points $\tau$ with a one-shot TCI.
    Since the tensor becomes larger by adding a new leg, the low-rank structure of the tensor might be lost. However, in the case studied here, the low-rank structure is retained, as in the case of the strong-coupling expansion \cite{TCI_strong_coupling}.
    Thus, we adopt the second method below.

    We generate the sampling points $\tau$ by using \texttt{SparseIR.jl} \cite{sparseIR1,sparseIR2,sparseIR3,sparseIR4} in the DMFT analysis presented later.
    It is known that the Matsubara Green's functions are highly compressible and the whole information of the Green's function can be reconstructed from a few sampling points based on the intermediate representation \cite{sparseIR1}.
    The library provides us with a routine to generate the sampling points and reconstruct the Green's function.
    This technique is useful to reduce the computational cost for the calculation of the Green's function in the TCI algorithm.

    \subsubsection{Change of variables}
    As in the case of the partition function, the integral in Eq.~\eqref{eq:Green's function integral} is defined over a simplex, which needs to be transformed into a hypercube.
    Furthermore, since $\tilde{\bm{D}}_{n}^{\sigma}$ depends on $\mathcal{G}_{\sigma}(\tau - \tau_{i})$ that has a discontinuous jump at $\tau = \tau_i$, the integrand $Q^{\sigma}(\tau_{1}, \cdots, \tau_{n}; \tau)$ is also discontinuous at $\tau_{i} = \tau$ for $1 \le i \le n$. 

    To eliminate the discontinuities, we divide the simplex $S_{n}^{0,\beta}$ into smaller $n+1$ regions ($S_k^{0,\tau}\times S_{n-k}^{\tau,\beta}$ with $k=0,1,\dots,n$), 
    \begin{multline}
        \mathcal{J}_{n}^{\sigma}(\tau) = (-U)^{n} \sum_{k=0}^{n} \int_{S_{k}^{0,\tau}} \dd{\tau_{1}} \cdots \dd{\tau_{k}} \\ \times\int_{S_{n-k}^{\tau, \beta}} \dd{\tau_{k+1}} \cdots \dd{\tau_{n}} 
        Q^{\sigma}(\tau_{1}, \cdots, \tau_{n}; \tau).
    \end{multline}
    In the region labeled by $k$, $\tau$ satisfies $\tau_{1} \le \cdots \le \tau_{k} \le \tau \le \tau_{k+1} \le \cdots \le \tau_{n}$, and thus, the integrand $Q^{\sigma}(\tau_{1}, \cdots, \tau_{n}; \tau)$ is continuous within each domain.

    To deform the integral domain to a hypercube, we perform the following variable transformation for each $k$:
    \begin{align}
        (\tau_{1}, \cdots, \tau_{k}) &= h_{k}^{0,\tau}(v_{1}, \cdots, v_{k}), \label{eq:Green's funcion change variable1} \\
        (\tau_{k+1}, \cdots, \tau_{n}) &= h_{n-k}^{\tau, \beta}(v_{k+1}, \cdots, v_{n}) \label{eq:Green's function change variable2}.
    \end{align}
    With these transformations, the integration domain changes from $S_{k}^{0,\tau} \times S_{n-k}^{\tau,\beta}$ to $[0,1]^{k} \times [0,1]^{n-k} = [0,1]^{n}$, and we get
    \begin{multline}
         \mathcal{J}_{n}^{\sigma}(\tau) = \sum_{k=0}^{n} (-U)^{n} \int_{[0,1]^{n}} \dd{v_{1}} \cdots \dd{v_{n}} Q_{k}^{\sigma}(v_{1}, \cdots, v_{n}; \tau) \\
         \times J_{h_{k}^{0,\tau}}(v_{1}, \cdots, v_{k}) J_{h_{n-k}^{\tau, \beta}}(v_{k+1}, \cdots, v_{n}).
    \end{multline}
    Here, $Q_{k}^{\sigma}(v_{1}, \cdots, v_{n}; \tau)$ is the function that satisfies
    \begin{equation}
        Q_{k}^{\sigma}(v_{1}, \cdots, v_{n}; \tau) = Q^{\sigma}(\tau_{1}, \cdots, \tau_{n}; \tau), 
    \end{equation}
     where $(\tau_{1}, \cdots, \tau_{n})$ and $(v_{1}, \cdots, v_{n})$ are related through Eqs.~\eqref{eq:Green's funcion change variable1} and \eqref{eq:Green's function change variable2}.
     As in the case of the partition function, the Jacobian is separable, i.e., it can be written in the form of
     \begin{multline}
         J_{h_{k}^{0,\tau}}(v_{1}, \cdots, v_{k}) J_{h^{\tau, \beta}_{n-k}}(v_{k+1}, \cdots, v_{n}) \\
         =B(\tau) B_{1}(v_{1}) \cdots B_{n}(v_{n}).
     \end{multline}
     
     For later convenience, we define a function,
     \begin{multline}
         \tilde{Q}_{k}^{\sigma}(v_{1}, \cdots, v_{n}; \tau) = Q_{k}^{\sigma}(v_{1}, \cdots, v_{n}; \tau) \\
         \times J_{h_{k}^{0,\tau}}(v_{1}, \cdots, v_{k}) J_{h_{n-k}^{\tau, \beta}}(v_{k+1}, \cdots, v_{n}), \label{eq:Green's function integrand without Jacobian}
     \end{multline}
     and write $\mathcal{J}_{n}^{\sigma}(\tau)$ as
     \begin{equation}
         \mathcal{J}_{n}^{\sigma}(\tau) = \sum_{k=0}^{n} (-U)^{n} \int_{[0,1]^{n}} \dd{v_{1}} \cdots \dd{v_{n}} \tilde{Q}_{k}^{\sigma}(v_{1}, \cdots, v_{n}; \tau). 
     \end{equation}
     With this, the Green's function can be expressed as
     \begin{multline}
         G_{\sigma}(\tau) \simeq \frac{Z_{0}}{Z} \sum_{n=1}^{n_{\mathrm{max}}} \sum_{k=0}^{n} (-U)^{n} \int_{[0,1]^{n}} \dd{v_{1}} \cdots \dd{v_{n}} \\ \times\tilde{Q}_{k}^{\sigma}(v_{1}, \cdots, v_{n}; \tau). \label{eq:Green integrand on the hypercube}
     \end{multline}

     \subsubsection{Discrete summation over \texorpdfstring{$k$}{Lg}}
    In addition to the integral over $v_{1}, \cdots, v_{n}$, there appears a discrete summation over $k$ in Eq.~\eqref{eq:Green integrand on the hypercube}. We have tried three ways to perform the discrete sum with respect to $k$.
    
    The first method is to get the tensor-train approximation of $\sum_{k=0}^{n} \tilde{Q}_{k}^{\sigma}(v_{1}, \cdots, v_{n}; \tau)$ by applying the TCI algorithm and integrate it over a hypercube by summing over the indices of the tensor train.
    We found that this method works up to the order of $n_{\mathrm{max}} \sim 20$, but it produces non-smooth Green's functions for higher orders.
    
    The second method treats $k$ as an additional leg of the tensor, together with $\tau_{1}, \cdots, \tau_{n}, \tau$, and applies the TCI algorithm to this extended tensor.
    By summing over the indices corresponding to $\tau_{1}, \cdots, \tau_{n}$ and $k$ in the obtained tensor train, we can perform the integral and discrete summation in Eq.~\eqref{eq:Green integrand on the hypercube} at once.
    Since the computational cost of evaluating $\tilde{Q}_{k}^{\sigma}(v_{1}, \cdots, v_{k}; \tau)$ is approximately $n+1$ times smaller than that of $\sum_{k=0}^{n} \tilde{Q}_{k}^{\sigma}(v_{1}, \cdots, v_{n}; \tau)$, the TCI algorithm runs faster in this implementation as compared to the first one.
    However, the bond dimension required to achieve an accurate tensor-train approximation can grow higher due to the additional leg corresponding to $k$.
    In fact, we observed that the Green's function calculated up to the 30th order by this methods was not smooth even with a maximum bond dimension $\chi = 400$.

    The third method is to find the tensor-train representation of $\tilde{Q}_{k}^{\sigma}(v_{1}, \cdots, v_{n})$ by the TCI algorithm and calculate $\int_{[0,1]^{n}} \dd{v_{1}} \cdots \dd{v_{n}} \tilde{Q}_{k}^{\sigma}(v_{1}, \cdots, v_{n})$ for each $k$ independently, and finally take a summation over $k$. 
    This method is more stable than the other methods since the integrand has a simpler structure. With this approach, we can obtain a smooth Green's function even if the bond dimension is taken to be $\chi = 50$.
    Therefore, we adopt the third method in the analysis presented in Sec.~\ref{sec:results}.

    As in the case of the partition function, no significant change in the efficiency is observed with the reordering of the legs, so that we use the most natural order $(v_{1}, \cdots, v_{n}, \tau)$ in the actual calculations.

\subsection{Rough estimate of the required maximum order} \label{subsec:rough estimation for n_max}
    Whenever we calculate the partition function or the Green's function for the impurity problem, we have to choose the maximum order $n_{\mathrm{max}}$ in such a way that the truncation does not change the results much.
    Here, we roughly estimate $n_{\mathrm{max}}$ required to obtain an accurate result. 
    
    Let us focus on the DMFT effective impurity problem for the Hubbard model in the Mott regime. In this parameter region, the local Green's function and hence the hybridization function quickly decays in imaginary time. Therefore, we can neglect $\Delta_\sigma(\tau)$ in Eq.~(\ref{eq:Weiss field}), and $\mathcal{G}_{\sigma}(\tau)$ is approximately given by a constant,
    \begin{equation}
        \mathcal{G}_{\sigma}(\tau) \approx -1/2\quad \text{for } 0 \le \tau \le \beta. \label{eq:approximated Weiss field}
    \end{equation}
    Although this approximation might be oversimplified, it captures the typical behavior of the Weiss field in the low-temperature and strong-coupling region.
    Within this approximation, the partition function can be analytically evaluated as
    \begin{equation}
        \frac{Z}{Z_{0}} = \sum_{n\text{:even}} \frac{1}{n!} \qty(\frac{\beta U}{4})^{n}.
    \end{equation}
    This means that the even-order contribution is proportional to the Poisson distribution with both the mean and the variance being equal to $\beta U / 4$.
    When $\beta U / 4 \gg 1$, the Poisson distribution can be further approximated by the Gaussian distribution.  From the property of the Gaussian distribution, we can see that if we take
    \begin{equation}
        n_{\mathrm{max}} \simeq \frac{\beta U}{4} + 3 \sqrt{\frac{\beta U}{4}},
        \label{eq:estimation of n_max}
    \end{equation}
    it can cover $3\sigma$ of the Gaussian distribution, which is sufficient to get an almost exact value of $Z/Z_{0}$.

    So far, we have focused on the low-temperature and strong-coupling regime.
    As we increase the temperature or decrease the interaction strength, $|\mathcal{G}_{\sigma}(\tau)|$ generally decreases from $1/2$, and this will shift the center of the distribution to a smaller value than $\beta U / 4$.
    Therefore, the estimate \eqref{eq:estimation of n_max} is also sufficient for the weak-coupling and high-temperature regime.

\section{Results} \label{sec:results}
  \subsection{Exactly solvable impurity model} \label{subsec:exactly solvable model result}
  We first benchmark our calculations with an exactly solvable impurity model, which corresponds to an impurity problem for the Falicov--Kimball model in DMFT \cite{FK_model}. The action is given by Eq.~\eqref{eq:SIAM action} with $\Delta_{\downarrow}(\tau) = 0$, which means that the spin-down impurity electron does not hybridize with the bath degrees of freedom. 
  In this case, we can easily integrate out the spin-down electrons for arbitrary $\Delta_{\uparrow}(\tau)$. The Green's function of the spin-up electron can be calculated analytically as
  \begin{equation}
      G_{\uparrow}(i\omega_{n}) = \frac{1}{2} \qty( \frac{1}{\mathcal{G}_{\uparrow}(i\omega_{n})^{-1} - U/2} + \frac{1}{\mathcal{G}_{\uparrow}(i\omega_{n})^{-1} + U/2} ),
  \end{equation}
  where 
  \begin{equation}
      \mathcal{G}_{\uparrow}(i\omega_{n}) = \frac{1}{i\omega_{n} - \Delta_{\uparrow}(i \omega_{n})}.
      \label{eq:Relation btw Weiss and hybridization}
  \end{equation}

 \begin{figure*}[t]
    \centering
    \includegraphics[width=0.9\linewidth]{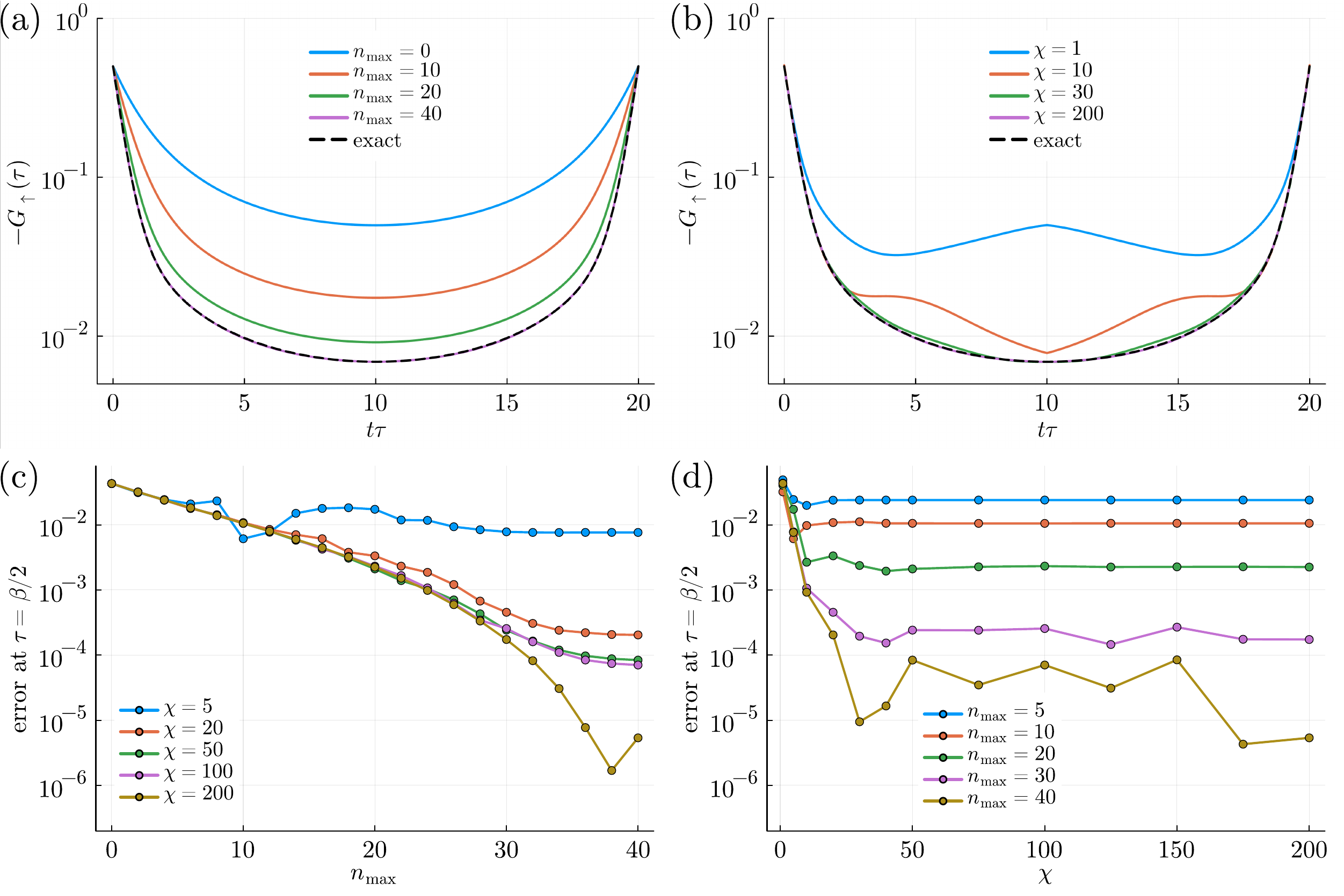}
    \caption{
    Results of the weak-coupling TCI solver for the exactly solvable impurity model with $\mathcal{G}_{\downarrow}(\tau) = -1/2$ and $\mathcal{G}_{\uparrow}(\tau)$ given by Eq.~(\ref{eq:Weiss field Bethe}).
    The parameters are set to $\beta = 20/t$, $U=5t$.
    [(a), (b)] The Green's functions obtained from the TCI solver and the exact solution for (a) several maximum orders $n_{\rm max}$ with the maximum bond dimension $\chi=200$, and (b) several $\chi$ with $n_{\rm max}=40$.
    [(c), (d)] The difference of the Green's functions at $\tau = \beta/2$ between the TCI and the exact solution plotted as a function of (c) $n_{\mathrm{max}}$ and (d) $\chi$.}
    \label{fig:FK model}
  \end{figure*}

  One of the simplest choices for $\Delta_{\uparrow}(\tau)$ is $\Delta_{\uparrow}(\tau) = 0$.
  In this case,  $\mathcal{G}_{\uparrow}(\tau) = \mathcal{G}_{\downarrow}(\tau) = -1/2$ for $0 \le \tau \le \beta$, and $P(h^{0,\beta}_{n}(v_{1}, \cdots, v_{n}))$ in Eq.~\eqref{eq:partition integrand without Jacobian} and $Q_{k}^{\sigma}(v_{1}, \cdots, v_{n})$ in Eq.~\eqref{eq:Green's function integrand without Jacobian} become constant. 
  As mentioned in Sec.~\ref{subsec:computation of partition function} and Sec.~\ref{subsec:computation of Green's function}, the Jacobians are separable, so that the entire integrands for the partition function and the Green's function are also separable in this case.
  Therefore, they trivially have tensor-train representations with rank $\chi = 1$, which is not suitable for the benchmark. 

  To benchmark the performance of the weak-coupling TCI impurity solver, we take $\mathcal{G}_{\uparrow}(\tau)$ to be the noninteracting Green's function on the Bethe lattice with infinite coordination number,
  \begin{equation}
      \mathcal{G}_{\uparrow}(\tau) = - \int_{-2t}^{2t} \dd{\omega} \frac{\sqrt{4t^2 - \omega^2}}{2 \pi t^2} \frac{e^{-\tau \omega}}{1 + e^{-\beta \omega}},
      \label{eq:Weiss field Bethe}
  \end{equation}
  which is a typical initial Green's function used in DMFT calculations.

  In Fig. \ref{fig:FK model}(a) and (b), we show the numerical results for the Green's function $G_{\uparrow}(\tau)$ with $\beta = 20/t$ and $U=5t$ calculated by the weak-coupling TCI impurity solver, which are compared with the exact solution. Figure \ref{fig:FK model}(a) shows how the result changes by increasing the maximum order $n_{\mathrm{max}}$ while keeping the maximum bond dimension fixed at $\chi = 200$.
  With $n_{\mathrm{max}} = 20$, there is a difference of around $10^{-3}$ as compared to the exact one, while the result with $n_{\mathrm{max}} = 40$ agrees with the exact solution almost perfectly.
  This is consistent with the rough estimate we discussed in Sec.~\ref{subsec:rough estimation for n_max}. 
  In Fig. \ref{fig:FK model}(b), the Green's function for several values of $\chi$ are plotted with $n_{\mathrm{max}}$ being fixed at $n_{\rm max}=40$.
  One can see that the tensor-train approximation of rank $\chi = 30$ gives a smooth result which almost agrees with the exact solution. 

  In Fig.~\ref{fig:FK model}(c) and (d), we present a more detailed analysis of the convergence of the results with respect to $n_{\rm max}$ and $\chi$. 
  In Fig.~\ref{fig:FK model}(c), we plot the deviation of the Green's function from the exact solution at $\tau = \beta/2$ as a function of $n_{\mathrm{max}}$. 
  Different curves correspond to different fixed values of $\chi$.
  For $\chi \le 100$, we can see a saturation behavior, i.e., the results are not improved anymore even if $n_{\rm max}$ is increased.
  Although evaluating the integral with $n_{\rm max}$ close to or larger than $40$ is computationally demanding with $\chi \le 100$, the error becomes smaller than $10^{-4}$, which is often enough for practical purposes.
  For $\chi=200$, we do not find an obvious sign of saturation up to $n_{\mathrm{max}}=40$, where the precision of $10^{-5}$ can be reached.
  
  Figure \ref{fig:FK model}(d) shows the deviation of the Green's function at $\tau = \beta / 2$ as a function of $\chi$ with several values of $n_{\mathrm{max}}$.
  For $n_{\mathrm{max}} \le 30$, the error decreases exponentially with respect to the bond dimension in the range of $0 \le \chi \le 15$. For $\chi > 50$, the error remains nearly constant.
  This indicates that the integral with $n_{\rm max}\le 30$ can be efficiently computed by the weak-coupling TCI.
  For $n_{\mathrm{max}} = 40$, the error does not exhibit a plateau over the range of $0 \le \chi \le 200$, but setting $\chi = 200$ results in an error of $10^{-5}$, which is sufficiently small.

  The fast convergence with respect to the bond dimension indicates that the integrand in the weak-coupling expansion has a low-rank structure that can be searched by the TCI algorithm.
  Since the Jacobians are separable as mentioned in Sec.~\ref{subsec:computation of partition function} and Sec.~\ref{subsec:computation of Green's function}, 
  whether the whole integrand has a low-rank structure or not depends on the structure of $P(h_{n}^{0,\beta}(v_{1}, \cdots, v_{n}))$ and $Q_{k}^{\sigma}(v_{1}, \cdots, v_{n})$ in Eqs.~\eqref{eq:partition integrand without Jacobian} and \eqref{eq:Green's function integrand without Jacobian}.
  In the low-temperature regime, $\mathcal{G}_{\uparrow}(\tau)$ exhibits a plateau over a wide range of $0 < \tau < \beta$.
  As a result, $P(h_{n}^{0,\beta}(v_{1}, \cdots, v_{n}))$ and $Q_{k}^{\sigma}(v_{1}, \cdots, v_{n})$ take almost constant values in most parts of the hypercube.
  Thus we can expect that the integrands are almost separable in the low temperature regime.
  This may be the reason why the TCI algorithm works well in the weak-coupling expansion.
  The effective impurity model appearing in DMFT also shares a similar property.
  In the zero-temperature or strong-coupling limit, the local Green's function $G_{\mathrm{loc}}(\tau)$ approaches zero in almost the entire interval $0<\tau<\beta$, and the Weiss field $\mathcal{G}_{\sigma}(\tau)$, which is related to $G_{\mathrm{loc}}(\tau)$ through Eqs.~\eqref{eq:Weiss field} and \eqref{eq:new hybridization function}, becomes almost constant over $0 < \tau < \beta$. 
  Thus, the integrand is almost separable in these limits, and it is expected to be so also in the finite but low-temperature and strong-coupling regimes.
    \begin{figure*}[t]
    \centering
    \includegraphics[width=0.9\linewidth]{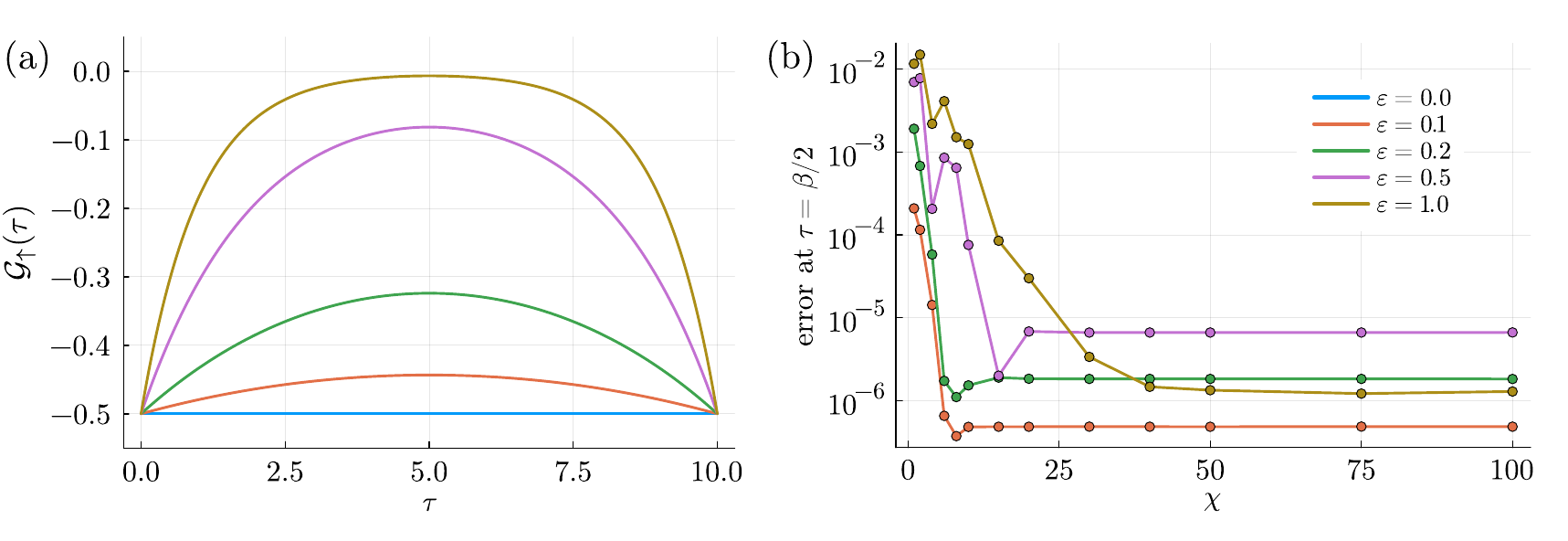}
    \caption{
    Results of the weak-coupling TCI solver for the exactly solvable impurity model with constant hybridization functions $\Delta_{\uparrow}(\tau) = -\varepsilon^2/2$ and $\Delta_{\downarrow}(\tau)=0$. The parameters are set to $\beta=10$, $U=5$.
    (a) The Weiss field $\mathcal{G}_{\uparrow}(\tau)$ [Eq.~\eqref{eq:Weiss field constant hyb}] corresponding to the hybridization function $\Delta_{\uparrow}(\tau) = -\varepsilon^2/2$ for several values of $\varepsilon$.
    (b) The difference of the Green's functions at $\tau = \beta/2$ between the TCI and the exact solution plotted as a function of $\chi$. The maximum perturbation order is fixed to $n_{\mathrm{max}} = 24$.
    Since the TCI algorithm converges with the bond dimension $\chi = 1$ when $\varepsilon = 0$, the data for $\varepsilon=0$ are omitted from the plot.}
    \label{fig:convergence Delta dependence}
    \end{figure*}
  Figure~\ref{fig:convergence Delta dependence} shows how the form of the hybridization function affects the efficiency of the TCI approach.
  We calculate the Green's function $G_{\uparrow}(\tau)$ for impurity problems with constant hybridization functions,
  \begin{equation}
      \Delta_{\uparrow}(\tau) = -\frac{\varepsilon^2}{2}, \quad
      \Delta_{\downarrow}(\tau) = 0, \label{eq:constant hyb}
  \end{equation}
  for $\beta=10$ and $U=5$, and investigate the convergence of the error with respect to the bond dimension for several values of $\varepsilon$.
  For the hybridization function $\Delta_{\uparrow}(\tau)$ in Eq.~\eqref{eq:constant hyb}, the Weiss field $\mathcal{G}_{\uparrow}(\tau)$ can be analytically calculated from Eq.~\eqref{eq:Relation btw Weiss and hybridization} as
  \begin{equation}
      \mathcal{G}_{\uparrow}(\tau) = -\frac{1}{2} \frac{e^{-\tau \varepsilon} + e^{-(\beta - \tau) \varepsilon}}{1 + e^{-\beta \varepsilon}},
      \label{eq:Weiss field constant hyb}
  \end{equation}
  which is plotted for several values of $\varepsilon$ in Fig.~\ref{fig:convergence Delta dependence}(a).
  Figure~\ref{fig:convergence Delta dependence}(b) shows the deviation of the Green's function $G_{\uparrow}(\tau)$ at $\tau = \beta/2$ from the exact one as a function of $\chi$.
  When $\varepsilon = 0$ (i.e., when the Weiss field is constant), the integrand is exactly separable, and the TCI algorithm converges with $\chi = 1$ (which is why the result is omitted in Fig.~\ref{fig:convergence Delta dependence}(b)).
  As $\varepsilon$ increases and the Weiss field deviates from a constant function, the bond dimension at which the saturation can be observed becomes larger.

  The computational cost of the algorithm can be estimated as follows.
  To obtain the $n$th-order contribution to the Green's function, we have to decompose $n+1$ different tensors with $n+1$ legs by TCI.
  Here, the $n+1$ different tensors correspond to the integrands for each $k$ in Eq.~\eqref{eq:Green integrand on the hypercube}.
  When decomposing one of them with a bond dimension $\chi$, we have to sample its elements $\order{\chi^2 d n}$ times \cite{TCI_library}, where $d$ is the dimension of the legs corresponding to the order of the GK quadrature in this work.
  Since the sampling of elements requires a calculation of a determinant of an $n \times n$ matrix, whose computational cost scales as $\order{n^3}$, the total cost for sampling becomes $\order{\chi^2 d n^4}$.
  Therefore, recalling the computation of the $n$th-order term requires the decomposition of $n+1$ tensors, the computational cost for evaluating the $n$th-order term is $\order{\chi^2 d n^{5}}$.
  To obtain the contribution from the first order to the $n_{\mathrm{max}}$th order, the overall computational cost amounts to $\order{\chi^2 d n_{\mathrm{max}}^{6}}$.
  Combining this scaling with the rough estimate \eqref{eq:estimation of n_max}, the cost can be expressed in terms of $\beta$ and $U$ as $\order{\chi^2 d (\beta U)^{6}}$.

  \subsection{Dynamical mean-field theory applications} \label{subsec:DMFT result}
\begin{figure}[t]
        \centering
        \resizebox{\columnwidth}{!}{%
        \begin{tikzpicture}
           \draw[thick, blue, densely dashed] (0, 0) -- (7.5, 7.5/100 * 50);
           \draw (3.3, 3.6) node[right, blue]{$\beta U/4 + 3\sqrt{\beta U / 4} = 40$};
        
           \draw[->, semithick, >=latex] (-0.5, 0) -- (7.5, 0);  
           \draw[->, semithick, >=latex] (0, -0.5) -- (0, 4.5);  
           \draw (0,0) node[below left]{O};
           \draw (3.75, -0.7) node{$U/t$};
           \draw (-1.25, 2.25) node{$k_{\mathrm{B}}T / t$};

           \foreach \i in {1, 2, 3, 4, 5, 6, 7} {
               \draw[semithick] (\i, 0.1) -- (\i, 0.0) node[below] {\i};  
           }
           \foreach \i/\label in {1/0.02, 2/0.04, 3/0.06, 4/0.08} {
             \draw[semithick] (0.1, \i) -- (0.0, \i) node[left] {\label};   
           }
        
           \draw[thick] (4.5, 2.75) to [out=-88,in=98] (4.72, 0.0);
           \draw[thick] (4.5, 2.75) to [out=-75, in=145] (5.8, 0.0);
        
           \draw (6.5, 0.8) node{insulator};
           \draw (2.5, 0.8) node{metal};
           \draw (4.7, 1.5) to [out=70, in=190] (5.3, 2.0) node[right]{coexistence region};
        \end{tikzpicture}
        }
        \caption{
        DMFT phase diagram for the Hubbard model on the Bethe lattice. The dashed line ($\beta U /4 + 3 \sqrt{\beta U /4}=40$) corresponds to a rough estimate of the boundary above which the weak-coupling TCI solver in the current setup can provide accurate results for the maximum order $n_{\mathrm{max}} \simeq 40$ and bond dimension $\chi\lesssim 200$.
        }
        \label{fig:phase diagram}
    \end{figure}
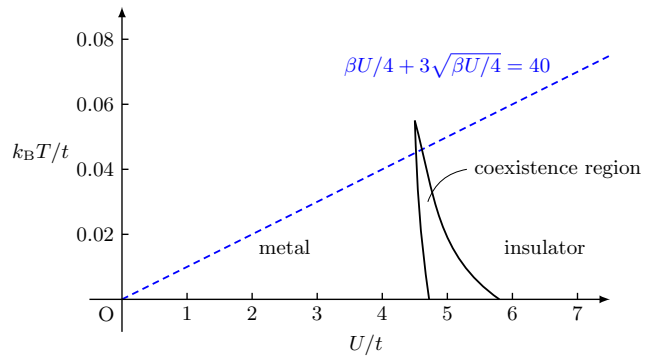

    Here we apply the weak-coupling TCI impurity solver to DMFT to study the Hubbard model on the Bethe lattice with infinite coordination number [Eq.~\eqref{eq:Hubbard model Bethe lattice}].
    We will show results for two inverse temperatures, $\beta = 16/t$ and $\beta = 20/t$.
    The critical endpoint of the Mott transition is known to lie in between these two temperatures \cite{MIT7} (see Fig.~\ref{fig:phase diagram}).
    Thus, we expect to observe a metal-insulator crossover at $\beta = 16/t$ and a Mott transition at $\beta = 20/t$.

    \subsubsection{Analysis of the crossover region}
    \begin{figure*}[t]
        \centering
        \includegraphics[width=\linewidth]{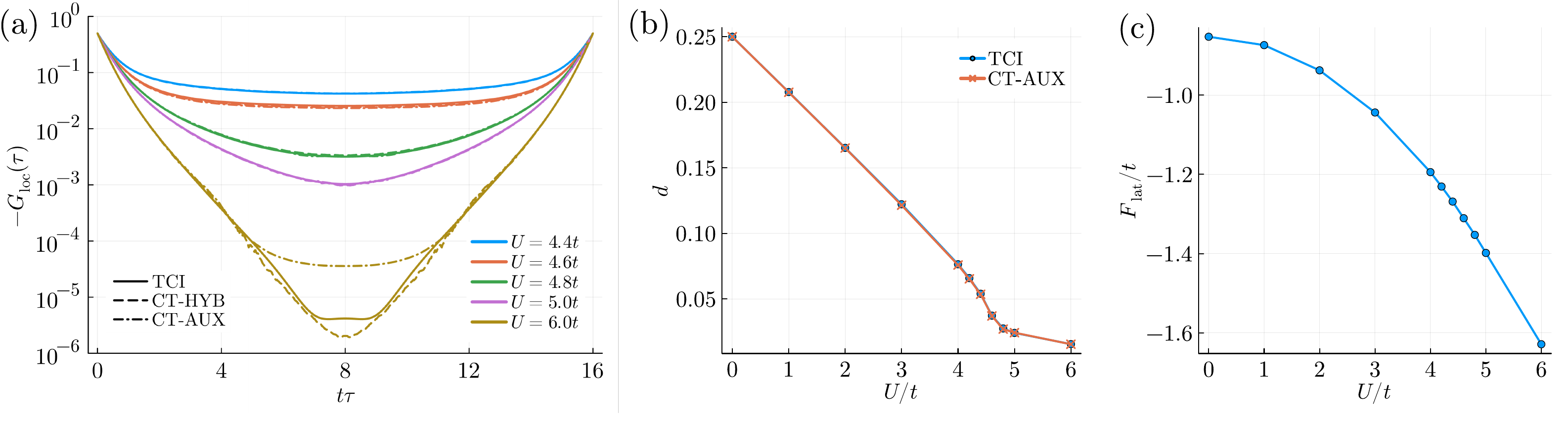}
        \caption{DMFT results for the half-filled Hubbard model on the Bethe lattice at $\beta = 16/t$.
        (a) Local Green's functions for several values of $U$ in the metal-insulator crossover region.
        The solid, dashed and dash-dotted lines correspond to the results of the weak-coupling TCI, CT-HYB, and CT-AUX impurity solvers, respectively.
        In the TCI calculations, the maximum bond dimension is set to $\chi = 200$ in the last few DMFT loops.
        The maximum order $n_{\mathrm{max}}$ is determined by the rough estimate \eqref{eq:estimation of n_max}.
        (b) Doublon number $d = \ev{n_{i\uparrow} n_{i\downarrow}}$ as a function of $U$ calculated by the weak-coupling TCI and CT-AUX impurity solvers.
        (c) Free energy of the lattice model, $F_{\mathrm{lat}}$, as a function of $U$ calculated by the weak-coupling TCI solver.}
        \label{fig:DMFT beta16}
    \end{figure*}

    Figure \ref{fig:DMFT beta16}(a)-(c) shows the results of the weak-coupling TCI solver for $\beta = 16/t$, which is slightly above the Mott transition endpoint.
    From the analysis of the exactly solvable model, we concluded that a bond dimension $\chi \simeq 50$ provides reasonably accurate results.
    Therefore, we first iterate the DMFT loop with $\chi = 50$ to achieve an approximate convergence.
    Subsequently, we increase the bond dimension to $\chi = 200$ to obtain final results with high precision.
    The maximum order $n_{\mathrm{max}}$ is determined from the rough estimate \eqref{eq:estimation of n_max}.
    
    In Fig.~\ref{fig:DMFT beta16}(a), we plot the local Green's functions calculated by the weak-coupling TCI solver, and the continuous-time quantum Monte Carlo methods based on the hybridization expansion (CT-HYB) \cite{strong_coupling_expansion} and the auxiliary field (CT-AUX) \cite{weak_coupling_expansion2} formulation. 
    As we increase the interaction strength $U$, $-G_{\mathrm{loc}}(\beta / 2)$, which is roughly proportional to the density of states at the Fermi level, approaches zero.
    This behavior reflects the metal-insulator crossover.
    In the parameter range shown in Fig.~\ref{fig:DMFT beta16}(a), the results of the TCI solver agree well with those of the QMC methods. The difference is on the order of $10^{-4}$ or even smaller, suggesting that the crossover region can be studied by the TCI solver with an accuracy comparable to that of the QMC methods. There are slight differences between CT-HYB and CT-AUX, which
    give an idea of the uncertainties associated with the Monte Carlo errors and possibly a lack of full convergence of the DMFT loop. The two Monte Carlo methods are based on different types of perturbative expansions (CT-HYB is based on the strong-coupling expansion, while CT-AUX is based on the weak-coupling expansion). In the crossover regime, 
    CT-HYB is more efficient than CT-AUX \cite{CTQMC_comparison}.

    In the TCI approach, we can evaluate the doublon number $d = \ev{n_{i\uparrow} n_{i\downarrow}}$, which captures the effect of strong correlations, by using the formula \cite{CT-QMC},
    \begin{equation}
        d = \frac{1}{4} - \frac{\ev{n}_{\mathrm{wc}}}{\beta U},
        \label{eq:doublon number}
    \end{equation}
    where $\ev{n}_{\mathrm{wc}}$ is the average order of the weak-coupling expansion of the partition function defined by
    \begin{equation}
        \ev{n}_{\mathrm{wc}} = \frac{Z_{0}}{Z} \sum_{n=0}^{\infty} n \mathcal{I}_{n}.
    \end{equation}
    Since the TCI solver evaluates $\mathcal{I}_{n}$ for each $n$, the doublon number can be readily extracted.
    The doublon numbers calculated by the weak-coupling TCI solver and the CT-AUX solver are shown in Fig.~\ref{fig:DMFT beta16}(b).
    The results are consistent with each other.
    The curve exhibits a different slope for interaction strengths below and above $U \simeq 4.7 t$, which corresponds to the metal-insulator crossover.

    The free energy, which is difficult to access by the CT-QMC solver, can be directly calculated by the TCI solver.
    The free energy of the effective impurity model is related to $Z/Z_{0}$ by
    \begin{equation}
        F_{\mathrm{imp}} 
        = - \frac{1}{\beta} \ln \qty(\frac{Z}{Z_{0}}) - \frac{2}{\beta} \ln \qty(2 \hspace{-2mm} \prod_{n=-\infty}^{\infty} \hspace{-1mm} \qty(1 - \frac{\Delta(i\omega_{n})}{i\omega_{n}})),
        \label{eq:free energy of impurity model}
    \end{equation}
    where the second term corresponds to the free energy of the noninteracting impurity model \cite{FK_model}.
    The free energy of the whole lattice model, $F_{\mathrm{lat}}$, is related to $F_{\mathrm{imp}}$ by \cite{free_energy1,free_energy2}
    \begin{equation}
        F_{\mathrm{lat}} = F_{\mathrm{imp}} - \frac{1}{\beta} \sum_{n=-\infty}^{\infty} \frac{\Delta(i \omega_{n})^2}{t^2}.
        \label{eq:free energy of lattice}
    \end{equation}
    The free energy $F_{\mathrm{lat}}$ calculated by Eqs.~\eqref{eq:free energy of impurity model} and \eqref{eq:free energy of lattice} is shown in Fig.~\ref{fig:DMFT beta16}(c). We can see that $F_{\rm lat}$ monotonically decreases with $U$.
    Since there is no Mott transition at $\beta = 16/t$, the free energy is a smooth function of $U$.

    \subsubsection{Analysis of the Mott transition}
    Next we show the results around the Mott transition in the Hubbard model on the Bethe lattice at $\beta = 20/t$.
    As the temperature is lowered, the required perturbation order $n_{\mathrm{max}}$ increases, making the calculation more challenging.
    Our weak-coupling TCI solver can explore the temperature regime slightly below the critical endpoint of the Mott transition (Fig.~\ref{fig:phase diagram}).

    \begin{figure*}
        \centering
        \includegraphics[width=\linewidth]{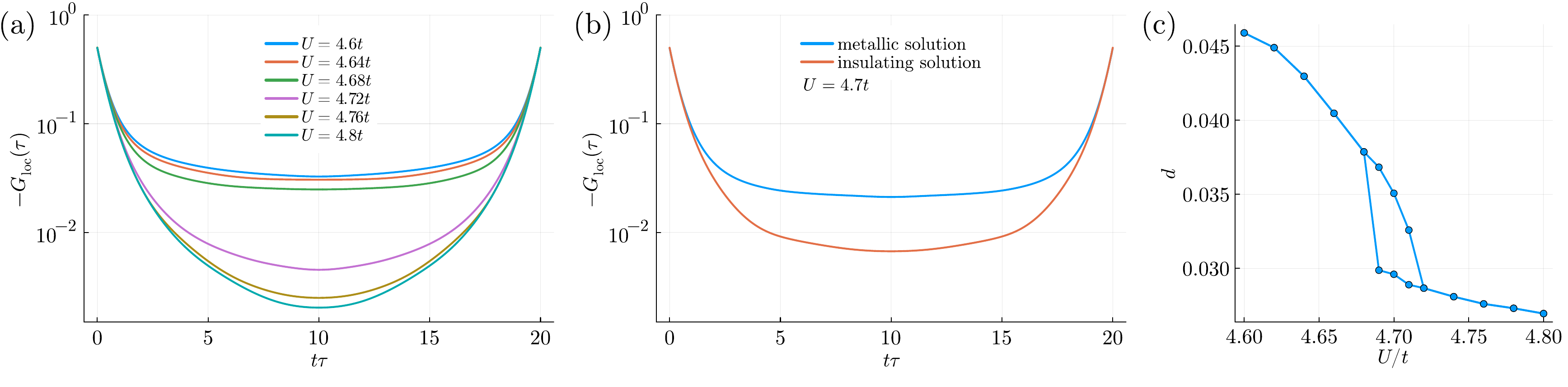}
        \caption{DMFT results obtained by the weak-coupling TCI impurity solver for the half-filled Hubbard model on the Bethe lattice at $\beta = 20/t$.
        (a) Local Green's function for several representative values of $U$. The value of the Green's function at $\tau = \beta/2$ suddenly drops at $U \simeq 4.7t$, which is an indication of the Mott transition.
        (b) The two coexisting DMFT solutions for $U = 4.7 t$. The blue line (metallic solution) is obtained by starting the DMFT loop with the solution for $U = 4.68t$, while the orange line (insulating solution) is obtained by starting with the solution for $U = 4.72t$.
        (c) Doublon number $d = \ev{n_{i\uparrow} n_{i\downarrow}}$ as a function of $U$. There is a hysteresis loop in the region $4.69 t \lesssim U \lesssim 4.71 t$.}
        \label{fig:DMFT beta20}
    \end{figure*}

    Figure \ref{fig:DMFT beta20}(a) shows the local Green's functions calculated by the TCI solver.
    One can observe that the value of the Green's function at $\tau = \beta/2$ suddenly drops when one slightly changes the interaction strength from $U = 4.68t$ to $U = 4.72t$.
    This discontinuous jump of $-G_{\mathrm{loc}}(\beta / 2)$ is an indication of the first-order Mott transition.
    
    As shown in Fig.~\ref{fig:DMFT beta20}(b), at $U = 4.7t$, the Green's function converges to two different solutions depending on the initial condition.
    Starting from the Green's function at $U = 4.68t$, it converges to the metallic solution depicted by the blue line in Fig.~\ref{fig:DMFT beta20}(b), while starting from the Green's function at $U = 4.72t$, it converges to the insulating solution depicted by the orange line.
    This coexistence of two solutions is a typical feature of a first-order phase transition.
    Even though we use the weak-coupling expansion, the coexistence behavior in the strong-coupling regime can be well captured by the TCI solver.

    The doublon number $d = \ev{n_{i\uparrow} n_{i\downarrow}}$ around the Mott transition is shown in Fig.~\ref{fig:DMFT beta20}(c).
    Since our TCI solver provides two stable solutions of the DMFT loop for $4.68t < U < 4.71t$, the doublon number exhibits a hysteresis loop in this region.

    \subsubsection{Current limitations of the weak-coupling TCI solver}
    To finish this section, let us discuss the limitations of the current weak-coupling TCI solver.
    From the analysis of the exactly solvable model and the comparison with CT-QMC in the DMFT calculations, we found that the weak-coupling TCI solver works well when the maximum order is $n_{\mathrm{max}} \lesssim 40$.
    However, for the low-temperature and strong-coupling region where $n_{\mathrm{max}} > 40$ is required, obtaining accurate results is difficult for our TCI solver, even with $\chi = 200$.
    This is because the integrals on the right hand side of Eq.~\eqref{eq:Green integrand on the hypercube} can be either positive or negative depending on $k$.
    In the Mott regime, the low-temperature Green's function takes very small negative values in a wide $\tau$ interval, and
    this is a consequence of the cancellation between positive and negative contributions in the summation over $k$.
    To evaluate the small negative values of the Green's function accurately, we need to calculate both positive and negative contributions of high-order terms with high precision, 
    and this requires a larger bond dimension $\chi$ and more computational time
    (this should be distinguished from the sign problem of QMC methods, since the issue may depend on the quantity to be calculated).
    
    In practice, we observed that the Green's function calculated by the TCI solver can become slightly positive when $\beta = 30/t, U=5t$ even though we set the bond dimension to $\chi = 200$, and this positive part survives even after several DMFT loops.
    This result violates the inequality $G_{\mathrm{loc}}(\tau) < 0 \, (0 < \tau < \beta)$, which should be satisfied by the physical Green's functions.

    In Fig.~\ref{fig:phase diagram}, we show a rough estimate of the boundary \eqref{eq:estimation of n_max} ($\beta U / 4 + 3 \sqrt{\beta U / 4} = 40$) at which the maximum order that we have to consider reaches $n_{\rm max}=40$. Above this line, our TCI solver works stably with  bond dimension $\chi\lesssim 200$, and the simulations can be run with small-scale parallelization. If one can handle larger bond dimensions, the boundary can be pushed down.

    In this work, we use the TCI algorithm which tries to minimize the error of each element of the tensor.
    The alternative choice is to use the one that minimizes the environment error, i.e., the error of the summation of all the elements \cite{TCI_noneq_weak_coupling, TCI_library}.
    This may improve the accuracy of the TCI solver, since we are interested in integrated values and not each individual element.
    Also, there could be an alternative variable transformation that may mitigate the positive and negative cancellation problem of the Green's functions.
    We will leave these issues as a future work.

\section{Discussions} \label{sec:discussions}
    We presented a weak-coupling tensor cross interpolation (TCI) solver for equilibrium quantum impurity problems.
    A major challenge in the weak-coupling expansion lies in evaluating high-order integrals, whose computational cost grows exponentially with their dimensions. 
    By employing the TCI algorithm, the integrand is decomposed into a product of matrix-valued functions, enabling an efficient evaluation of high-dimensional integrals through one-dimensional ones, which scales polynomially with respect to the dimension.
    With a suitable variable transformation, the integrand becomes a continuous function over a hypercube with a low-rank structure, enabling an efficient calculation.

    We have benchmarked the performance of our TCI solver for two setups: an exactly solvable impurity model and the DMFT solution of the Hubbard model with maximum perturbation order up to $n_{\rm max}=40$.
    For the exactly solvable model, we find that the error of the Green's function at $\tau = \beta / 2$
    decays exponentially as a function of $n_{\rm max}$ and the bond dimension $\chi$, and saturates at some point.
    An error on the order of $10^{-4}$ can be reached with a low-rank representation such as $\chi = 50$.
    For the DMFT calculation in the crossover region, we confirmed that the weak-coupling TCI solver can correctly reproduce the Green's functions obtained by QMC solvers.
    Using results obtained during the calculation of the Green's functions, we can also evaluate the doublon number and free energy of the lattice model with almost no extra effort. The latter quantity is difficult to calculate with QMC solvers.
    We also applied the TCI solver at lower temperature, where a first order Mott transition appears. 
    We demonstrated the coexistence of two stable solutions and the associated hysteresis behavior in the doublon number, which are characteristics of the metal-to-Mott insulator transition.
    These findings establish the weak-coupling TCI solver as a reliable and efficient 
    tool for the study of quantum impurity problems in the weak- and intermediate-interaction regimes. 

    The comparison of the TCI solver with other established methods highlights its strengths and limitations.
    One of the most notable advantages of the TCI solver over the CT-QMC methods is that it is free from the conventional sign problem. 
    While the models analyzed in this paper do not exhibit a sign problem in QMC, 
    the TCI solver will become more advantageous when applied to multi-orbital systems with off-diagonal hybridizations, multi-site clusters, systems with retarded spin interactions, spin-orbit coupled models or nonequilibrium impurity problems.
    In addition, the TCI solver facilitates the direct calculation of the free energy, which is a non-trivial problem for Monte Carlo techniques that lack information on the partition function.
    
    On the other hand, there is room to improve the efficiency of the TCI solver.
    The TCI algorithm requires a large number of evaluations of the determinant in Eqs.~\eqref{eq:expansion of partition function finite order} and \eqref{eq:expansion of Green's function finite order} to construct a tensor-train representation.
    Since the computational cost of calculating determinants scales as $\order{n^3}$ with the size $n$ of a matrix, the TCI algorithm becomes time-consuming for high-dimensional integrals.
    In CT-QMC, one evaluates the same type of determinants, but with a smaller cost of $\order{n^2}$ using the fast update algorithm \cite{CT-QMC}.
    In the TCI solver, such a technique is not available since the variable transformation changes the structure of the determinant. We will leave this issue as an interesting future problem.
    
    Compared with wave-function-based solvers such as exact diagonalization (ED) \cite{ED1, ED2, ED3} or the density matrix renormalization group (DMRG) \cite{DMRG1,DMRG2,DMRG3}, 
    the TCI solver has the advantages of being free from finite-size effects and easily applicable to calculations at nonzero temperatures. 
    Another method worth mentioning here is the influential functional (IF) method \cite{IF1,IF2,IF3,IF4,IF5}, a recently developed powerful solver for impurity problems. 
    The IF method is similar to the TCI solver in that it decomposes a tensor into a product of small matrices. 
    However, the objects being decomposed differ significantly: the TCI solver targets the integrand appearing in the perturbative expansion, while the IF method targets the discretized Feynman--Vernon IF, which encodes the time non-locality introduced by the bath.
    In the IF method, once the tensor-train representation of a given bath's IF is obtained, calculations for arbitrary interaction with that bath can be performed easily.
    Nevertheless, constructing the tensor-train representation for the IF is computationally demanding.
    Moreover, since the bath is self-consistently determined and hence is updated for each iteration in the case of DMFT, the tensor-train decomposition of the IF must be performed multiple times, making the calculation more time-consuming.
    Thus, when dealing with impurity problems or DMFT for small (large) $U$, the weak-coupling (strong-coupling) TCI solver should be more efficient.
    
    Finally, compared to the strong-coupling TCI solver \cite{TCI_strong_coupling}, the weak-coupling TCI solver is expected to be better suited for addressing multi-site cluster impurity problems.
    In the strong-coupling expansion, the integrand includes a trace over the impurity Hilbert space.
    If the impurity Hamiltonian is diagonal in the occupation number basis, the segment formalism \cite{CT-QMC} enables a fast calculation of the trace.
    However, if it is not diagonal, one has to perform multiplications of matrices whose size grows exponentially with the number of orbitals or sites, which becomes the bottleneck of the calculation.
    On the other hand, in the weak-coupling expansion, in addition to the integral over the imaginary time as in Eqs.~\eqref{eq:expansion of partition function} and \eqref{eq:expansion of Green's function}, we have discrete summations over the site indices in multi-site clusters or orbitals and interaction-type indices in multi-orbital systems.
    If the integrands with additional indices also exhibit a low-rank structure, similar to the single-site and single-orbital case, TCI can be used to efficiently evaluate the summations over these indices.
    In the weak-coupling TCI approach, the computational cost grows polynomially with the system size, the number of orbitals, and the number of interaction types.
    For multi-orbital models with general interactions, the number of interaction types grows as $\order{n_{\mathrm{orb}}^4}$, where $n_{\mathrm{orb}}$ is the number of orbitals, and the tensor to be decomposed becomes large. 
    Thus, treating such a multi-orbital model with general interactions will be challenging.
    It has been pointed out in Ref.~\cite{multi_orbital_QTCI} that the low-order skeleton diagrams in multi-orbital systems with retarded interactions exhibit a low-rank structure in the quantics representation~\cite{quantics1,quantics2}, which indicates the possibility of treating multi-orbital problems with the weak-coupling TCI.
    Whether higher-order diagrams also have a low-rank structure remains an open question, which should be investigated in future studies.
    
    For a single-site, single-orbital problem, the strong-coupling TCI approach exhibits a lower average perturbation order than the weak-coupling one over a wide range of $U$ \cite{CTQMC_comparison}, suggesting that the weak-coupling approach is more efficient only for small $U$ $(U/t \lesssim 1)$.
    Still, if the integrand of the weak-coupling expansion can be decomposed with lower bond dimensions than that of the strong-coupling expansion, the weak-coupling TCI approach could be preferable even for larger $U$.
    The detailed examination of this point is also left for a future study.
    
    Overall, while the current implementation of the TCI solver leaves room for refinements, its unique features and capabilities make it a promising addition to the toolkit for studying strongly correlated electron systems.
    The potential of the weak-coupling TCI solver will be more clearly demonstrated in future applications to problems where the CT-QMC methods suffer from the sign problem.

\section*{Acknowledgements} \label{sec:acknowledgements}
    S.M. and N.T. acknowledge support by JST FOREST (Grant No.~JPMJFR2131) and JSPS KAKENHI (Grant No.~JP24H00191).
    S.M. is also supported by the Forefront Physics and Mathematics Program to Drive Transformation (FoPM), a World-Leading Innovative Graduate Study (WINGS) Program, the University of Tokyo.  
    H.S. is supported by JSPS KAKENHI (Grants Nos.~21H01041, 21H01003, 22KK0226, and 23H03817) as well as JST FOREST (Grant No.~JPMJFR2232) and JST PRESTO (Grant No.~JPMJPR2012).
    P.W. is supported by SNSF Grant No.~200021-196966.
    The calculation has been done with a code based on \texttt{TensorCrossInterpolation.jl} \cite{TCI_library} and \texttt{SparseIR.jl} \cite{sparseIR4}.

\appendix
\section{Change of variables} \label{app:change var}
    In this Appendix, we review the variable transformations which are used to change the integral domain from the simplex to the hypercube and to avoid the discontinuities of the integrand.

\subsection{Transformation from \texorpdfstring{$S_{n}^{a,b}$}{Lg} to \texorpdfstring{$S_{n}^{0,1}$}{Lg}}
    Let us consider changing the integral domain from $S_{n}^{a,b}$ into $S_{n}^{0,1}$ through a variable transformation.
    To this end, we use a map $f_{n}^{a,b} : S_{n}^{0,1} \rightarrow S_{n}^{a,b}$, 
    \begin{equation}
        [f_{n}^{a,b}(\bm{y})]_{i} = a + (b-a)y_{i},
        \label{eq:map from normalized simplex to simplex}
    \end{equation}
    where $\bm{y} = (y_{1}, \cdots, y_{n}) \in S_{n}^{0,1}$.
    This map is clearly a bijection from $S_{n}^{0,1}$ to $S_{n}^{a,b}$ and its Jacobian is 
    \begin{equation}
        J_{f_{n}^{a,b}}(\bm{y}) = \det \qty(\pdv{f_{n}^{a,b}}{y}) = (b-a)^{n}.
    \end{equation}
    If we relate the variables $\bm{x} \in S_{n}^{a,b}$ and $\bm{y} \in S_{n}^{0,1}$ by
    \begin{equation}
        \bm{x} = f_{n}^{a,b}(\bm{y}), 
    \end{equation}
    the integral over $S_{n}^{a,b}$ with respect to $\bm{x}$ can be rewritten as an integral over $S_{n}^{0,1}$ with respect to $\bm{y}$:
    \begin{equation}
        \int_{S_{n}^{a,b}} \dd[n]{\bm{x}} F(\bm{x}) = \int_{S_{n}^{0,1}} \dd[n]{\bm{y}} F(f_{n}^{a,b}(\bm{y})) (b-a)^n.
    \end{equation}\\

\subsection{Transformation from \texorpdfstring{$S_{n}^{0,1}$}{Lg} to \texorpdfstring{$[0,1]^{n}$}{Lg}}
    Let us consider changing the integral domain from the simplex $S_{n}^{0,1}$ to the unit hypercube $[0,1]^{n}$. 
    To this end, we use a map $g_{n} : [0,1]^{n} \rightarrow S_{n}^{0,1}$, 
    \begin{equation}
        [g_{n}(\bm{z})]_{i} = 1 - \prod_{j=1}^{i-1} (1-z_{j}),
        \label{eq:map from hypercube to normalized simplex}
    \end{equation}
    where $\bm{z} = (z_{1}, \cdots, z_{n}) \in [0,1]^{n}$.
    This transformation is essentially the same as the one used in Ref.~\cite{TCI_strong_coupling}.
    After some considerations, it becomes clear that this map is a bijection from $[0,1]^{n}$ to $S_{n}^{0,1}$.
    Since the Jacobi matrix of this map is a lower triangular matrix, the Jacobian can be easily calculated as
    \begin{equation}
        \begin{aligned}
            J_{g_{n}}(\bm{z}) &= \det \qty(\pdv{g_{n}}{z}) \\
            &= (1-z_{1})^{n-1} (1-z_{2})^{n-2} \cdots (1-z_{n-1}).
        \end{aligned}
    \end{equation}
    If we relate the variable $\bm{y} \in S_{n}^{0,1}$ with $\bm{z} \in [0,1]^{n}$ by 
    \begin{equation}
        \bm{y} = g_{n}(\bm{z}),
    \end{equation}
    the integral over $S_{n}^{0,1}$ with respect to $\bm{y}$ can be rewritten as an integral over $[0,1]^{n}$ with respect to $\bm{z}$:
    \begin{equation}
        \int_{S_{n}^{0,1}} \dd[n]{\bm{y}} F(\bm{y}) = \int_{[0,1]^{n}} \dd[n]{\bm{z}} F(g_{n}(\bm{z})) J_{g_{n}}(\bm{z}).
    \end{equation}

\subsection{Transformation from \texorpdfstring{$S_{n}^{a,b}$}{Lg} to \texorpdfstring{$[0,1]^n$}{Lg}}
    By combining the two maps, $f_{n}^{a,b}$ and $g_{n}$, we can construct a bijective map from $[0,1]^{n}$ to $S_{n}^{a,b}$,
    \begin{equation}
        h_{n}^{a,b} = f_{n}^{a,b} \circ g_{n}.
        \label{eq:map from hypercube to simplex}
    \end{equation}
    The Jacobian of this map is given by
    \begin{equation}
        \begin{aligned}
            J_{h_{n}^{a,b}}(\bm{z}) &= J_{f_{n}^{a,b}}(g_{n}(\bm{z})) J_{g_{n}}(\bm{z}) \\
            &= (b-a)^{n} (1-z_{1})^{n-1} \cdots (1 - z_{n-1}).
            \label{eq:Jacobian}
        \end{aligned}
    \end{equation}
    If we relate the variable $\bm{x} \in S_{n}^{a,b}$ with $\bm{z} \in [0,1]^{n}$ by
    \begin{equation}
        \bm{x} = h_{n}^{a,b}(\bm{z}),
    \end{equation}
    the integral over $S_{n}^{a,b}$ with respect to $\bm{x}$ can be rewritten as an integral over $[0,1]^{n}$ with respect to $\bm{z}$:
    \begin{equation}
        \int_{S_{n}^{a,b}} \dd[n]{\bm{x}} F(\bm{x}) = \int_{[0,1]^{n}} \dd[n]{\bm{z}} F(h_{n}^{a,b}(\bm{z}))J_{h_{n}^{a,b}}(\bm{z}).
    \end{equation}

\end{document}